\documentclass[lettersize,journal]{IEEEtran}
\usepackage{amsmath,amsfonts}
\usepackage{amssymb}
\usepackage{algorithm}
\usepackage{array}
\usepackage{textcomp}
\usepackage{stfloats}
\usepackage{caption}
\usepackage{subfigure}
\usepackage{url}
\usepackage[noend]{algpseudocode}
\usepackage{amsfonts}
\usepackage{color}
\usepackage{diagbox}
\usepackage{multirow}
\usepackage{booktabs}
\usepackage{verbatim}
\usepackage{hyperref}
\usepackage{verbatim}
\usepackage{graphicx}
\usepackage{cite}
\usepackage{enumerate}

\allowdisplaybreaks[4]
\begin{document}
\bibliographystyle{IEEEtran}
\title{Near-Field Channel Estimation for Extremely Large-Scale Circular RIS-Aided mmWave MIMO-NOMA System with Beam Squint Effect}

\author{
	Wanyuan~Cai,
	~Shunli~Hong,
	~Youming~Li,
	~Menglei~Sheng,
	~and~Mingjun~Huang
	\thanks{Manuscript received Month Day, 2026; revised Month Day, 2026. This work was supported in part by the National Key Research and Development Program of the Ministry of Science and Technology under Grant 2023YFC2809400 and in part by the National Natural Science Foundation of China under Grant 61571250. (Corresponding author: Youming Li.)}
	
	\thanks{Wanyuan~Cai, Shunli~Hong, Youming~Li, and Mingjun~Huang are with the Faculty of Electrical Engineering and Computer Science, Ningbo University, Ningbo, 315211, China (email: {2301100041@nbu.edu.cn}; {hsl@zjvtit.edu.cn}; {liyouming@nbu.edu.cn}; {2540460495@qq.com};).}
	\thanks{Menglei~Sheng is with the Production Research Center, EcoFlow Inc., Shenzhen 518000, China (email: {1298918854@qq.com}).}

}

\markboth{}%
{Shell \MakeLowercase{\textit{et al.}}: A Sample Article Using IEEEtran.cls for IEEE Journals}


\maketitle

\begin{abstract}
Near-field channel estimation under beam squint effect is critical to future 6G millimeter-wave (mmWave) systems equipped with reconfigurable intelligent surfaces (RIS). In this paper, firstly, we design an extremely large-scale circular RIS (XL-CRIS) architecture to construct an angle-invariant near-field region for MIMO-NOMA system, which can maintain a constant effective aperture, allowing for a unified channel modeling framework. Then, to enable efficient parameter extraction, we model the received wideband MIMO-NOMA signal as a third-order tensor which is used to develop a multi-stage channel estimation framework. Accordingly, we decompose the multi-variable problem into several low-dimensional sub-problems, while naturally preserving path-wise parameter pairing through the shared permutation matrix. Finally, we derive a vector-form Cramér–Rao bound (CRB) as a theoretical performance benchmark. To illustrate the effectiveness of the proposed method, numerical experiments are carried out and compared with the discussed methods.
\end{abstract}

\begin{IEEEkeywords}
Near-field channel estimation, extremely large-scale circular RIS, MIMO-NOMA system, beam squint effect, third-order tensor modeling.
\end{IEEEkeywords}

\section{Introduction}
With the advent of emerging technologies such as digital twins \cite{10742568}, augmented reality and virtual reality \cite{9040264}, and holographic communication \cite{10054381} poses challenges to existing 5G systems due to their high demand for transmission rates and spectral efficiency. Multiple-input multiple-output non-orthogonal multiple access (MIMO-NOMA) at millimeter-wave (mmWave) band has emerged as a promising candidate technology to meet these stringent requirements, as it synergizes the abundant spectrum resources of mmWave with the high spectral efficiency inherent in NOMA schemes and desirable beamforming gain of MIMO technology \cite{9693417}. Nevertheless, due to the naturally poor penetrability and substantial signal attenuation, mmWave signal is highly vulnerable to obstruction in line-of-sight (LoS) path, which often leads to extensive communication blind spots \cite{9847080}.

Fortunately, the reconfigurable intelligent surface (RIS) has risen as a potent remedy to overcome LoS blockages by establishing supplementary transmission links, thus eliminating the bottlenecks that limit system performance \cite{10858311}. Specifically, underpinned by metasurfaces consisted of sub-wavelength unit cells, RIS is capable of tailoring its electromagnetic responses, such as frequency, phase, amplitude, and polarization, to realize the required beamforming gain in a real-time reconfigurable manner during the light-matter interaction \cite{10858311} \cite{9206044}. Consequently, the development of RIS-aided mmWave MIMO-NOMA technology is one of the powerful tools to meet future communication requirements. However, the successful implementation of MIMO beamforming \cite{11030617}, optimal RIS control \cite{10858764}, as well as system data demodulation \cite{10925440} requires reliable channel state information (CSI). Moreover, the information hidden in CSI, such as angle and time delay, can be exploited to enable highly accurate localization of user equipment (UE), which is also a critical requirement for 6G communication \cite{10475369} \cite{9540372}. Thus, reliable CSI acquisition has become one of the most critical issues for RIS-aided mmWave MIMO-NOMA systems.

In recent years, various far-field channel estimation methods based on compressed sensing (CS) theory, such as orthogonal matching pursuit (OMP)-based method \cite{10053657}, atomic norm minimization (ANM)-based methods \cite{9398559} \cite{10630591}, and approximate message passing-based method \cite{10360256}, have been proposed for RIS-aided mmWave MIMO systems via exploiting the inherent sparse characteristics of mmWave channel. Apart from CS-based methods, a number of methods, such as canonical polyadic decomposition (CPD)-based methods \cite{10552118,10772118,CAI2026110466}, use the low-rank characteristics of the signal to perform dimensionality reduction decomposition by modeling the received signal as a high-dimensional tensor, achieving excellent channel estimation performance. However, with the evolution of communication systems, the number of array antennas or RIS elements will be deployed towards large-scale, usually causing a rapid expansion of the near-field region beyond the coverage area of a single base station (BS) serving at mmWave band \cite{9903389} \cite{9693928}. Therefore, the far-field channel model considered previously becomes inappropriate, as the wavefront cannot be simply approximated by a planar form. In contrast, a spherical wavefront model is required for accurate characterization, whose corresponding channel model introduces extra distance term to assist in characterizing the resultant energy spread effect \cite{9693928} \cite{10464973}. However, the introducing of extra distance term in near-field channel model destroys the sparse characteristics of channel in traditional angular-domain, increasing the complexity of channel estimation. Nevertheless, this near-field effect also brings many benefits, such as focusing signal power on specific angles and distances \cite{9738442} and increasing the spatial degree of freedom for communications \cite{10243590}. Thus, there is a critical need to design novel channel estimation methods specifically tailored for near-field scenarios, which has become a crucial research challenge. 

Research on RIS-aided near-field channel estimation is still in its infancy, yet it has exhibited remarkable potential. Although the angular sparsity is undermined, the near-field channel exhibits significant sparsity in the polar domain (angular-distance domain). By exploiting this polar-domain sparsity, the channel estimation can also be reformulated within a CS framework. In \cite{10081022} and \cite{10896729}, 3D distributed CS framework and 3D-polar-domain simultaneous orthogonal matching pursuit method are proposed for solving CS-based channel geometric parameters estimation problem, respectively. However, these methods encounter the grid-mismatch problem. Considering this, the expectation-maximization-based method and off-grid fast sparse Bayesian learning framework are proposed to solve the constructed sparse recovery problem in \cite{10707360} and \cite{10153711}, respectively. Apart from sparsity, the low-rank characteristics of the channel are exploited for robust channel estimation without grid. In \cite{10663714}, by leveraging the low-rank factorization of the effective channels, an unified collaborative low-rank approximation (CLRA)-based channel estimation method is proposed, suitable for deployment in both far-field and near-field scenarios. In order to address the limitations of estimators that falter under the high-rank conditions of mixed LoS/NLoS near-field channels, based on \cite{10663714},
a piece-wise CLRA framework is proposed in \cite{10907863}, which decomposes the intractable high-rank effective channel into low-rank substructures via RIS surface partitioning, significantly mitigating training overhead while maintaining superior estimation accuracy. However, these methods just consider narrowband systems. 

With the deployment of mmWave technology, the bandwidth of the system increases, making wideband signals more common, which will cause a well-known spatial beam squint effect in phased array antennas \cite{10271123} \cite{10500431}. Due to the variation of the incident angle of the incident wave with frequency, the beam squint effect will cause the array response to become strictly frequency-dependent in wideband systems. This phenomenon destroys the conventional low-rank structure of the channel in the frequency dimension, rendering the received signal tensor strictly high-rank. Accordingly, the state-of-the-art tensor modeling methods, such as methods in \cite{10552118,10772118,CAI2026110466}, which typically rely on the assumption of frequency-independent spatial signatures, suffer from significant model mismatch and performance degradation. To address the system performance degradation of the wideband system posed by beam squint effect, researchers have proposed various methods for RIS-aided wideband channel estimation, which can be mainly categorized under the framework of CS theory. In \cite{8882325}, by representing the channel as a function of physical parameters, a super-resolution CS method is proposed that can adaptively update its dictionary. In \cite{9409636}, a twin-stage OMP algorithm is proposed to determine the equivalent angles by searching for the peak correlation between spatial steering vectors and the reflected channel. Then, an optimized pilot scheme is developed to improve the channel parameters estimation performance via leveraging the cross-entropy theory. In \cite{10639461}, by dividing the procedure into slow-time and fast-time parameter estimation modes, a beam-squint-aware OMP method and a subspace-aware least squares are proposed for parameters estimation in two modes, respectively. In \cite{11146848}, an adaptive codebook is designed and a two-region super-resolution angle estimation (TR-SRAE) scheme is proposed, in which TR-SRAE scheme is extended for angle estimation in the central region, while the angle estimation is formulated as a maximum-posteriori problem solved by the developed Bayesian inference method. However, these methods does not consider near-field communication scenarios and require designing a specific basis, which can incur high computational complexity when the basis involves many variables.

Motivated by these, in this paper, we design extremely large-scale circular RIS (XL-CRIS) to create near-field propagation environments to enable more UEs benefit from near-field communication. By efficiently exploiting the rotational symmetry of uniform circular array (UCA) architecture of the XL-CRIS, an angle-invariant near-field region is generated. This property ensures that the effective array aperture of the UCA remains constant for all incident angles, thereby avoiding the near-field shrinkage issue encountered by the uniform linear array (ULA) at large angles of incidence \cite{10243590} \cite{10934779}. Thus, by appropriately designing aperture of XL-CRIS, communication can be within the near-field region, avoiding the complexity caused by mixing far-field and near-field communication. Then, through formulating the received signal as a third-order tensor, we propose a multi-stage channel estimation method that leverages the subspace orthogonality and correlation between parameters, which can avoid the problem of mismatched parameters in different paths. Notably, while this paper shares a similar foundational architecture with our previous work \cite{CAI2026110466}, it fundamentally differs in its propagation model (near-field rather than far-field), channel model (wideband instead of narrowband), and proposed solving method. The primary contributions of this paper are as follows:

(1) We establish an XL-CRIS-aided mmWave MIMO-NOMA near-field wideband system, considering the beam squint effect. In the system, an uniform near-field region is generated by designing the XL-CRIS, which extends the near-field coverage to the entire cellular cell, allowing for a unified channel modeling framework that bypasses the complex distinction between near-field and far-field users. 

(2) We model the received signal as a third-order tensor, and then propose a multi-stage channel estimation method. In the proposed method, we divide the channel parameters estimation problem into multiple subproblems, avoiding high-dimensional basis caused by involving multiple variables. It is worthy mentioning that by tensor modeling, the permutation matrix is shared, i.e. the channel parameters associated with each path are paired automatically, avoiding parameters matching problem. 

(3) A concise vector-form of Cramér-Rao bound (CRB) for channel parameters is derived to serve as a benchmark for comparison with the proposed method. Numerical results are conducted to validate the effectiveness of the proposed method.

\section{System Model}\label{Sec_System_Model}
\begin{figure}[htbp]
	\centering
	\includegraphics[width=3 in]{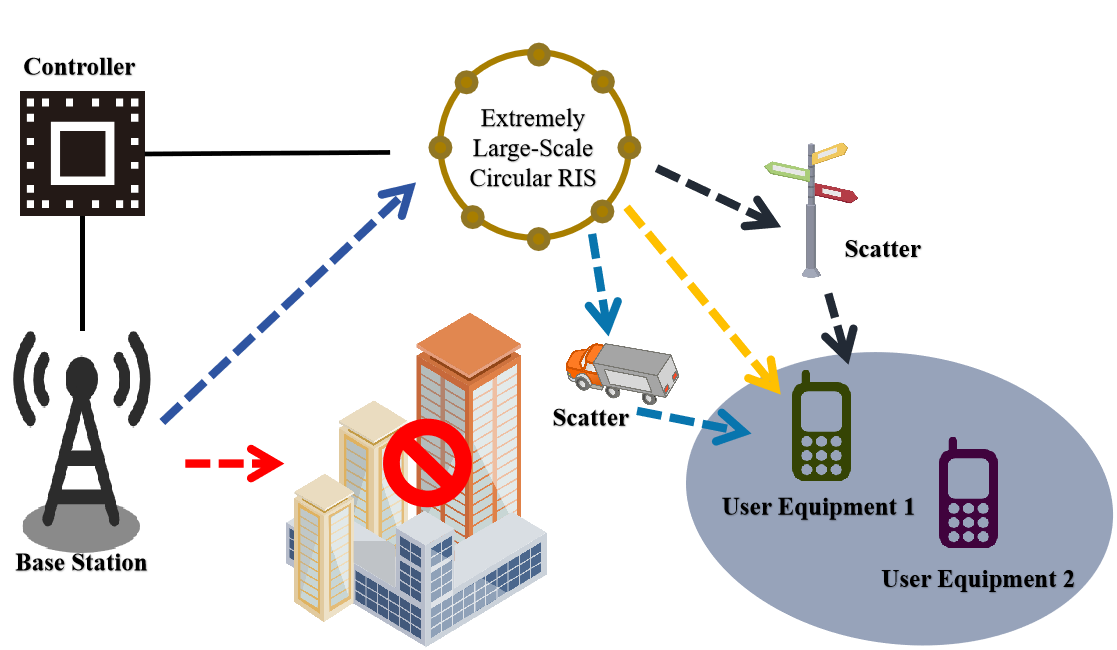}
	\caption{System model.}
	\label{System_pitcture}
\end{figure}

As shown in Fig. \ref{System_pitcture}, we consider a downlink XL-CRIS-aided mmWave MIMO-NOMA near-field wideband system, where an XL-CRIS is employed to establish extra communication links to overcome the blockage of the direct BS-UE LoS path caused by buildings. The designing of XL-CRIS reconstructs the channel, making traditional far-field channel models no longer applicable and necessitating a shift toward accurate near-field channel modeling. In the system, a half-wavelength spaced ULA is deployed at the BS and UE, with $N_{\rm BS}$ and $N_{\rm UE}$ antennas, respectively. To achieve hardware efficiency, the system incorporates a hybrid analog-digital design. This allows for the simultaneous transmission of $N_s$ streams while using a reduced number of RF chains ($M_{\rm BS}$ at the BS and $M_{\rm UE}$ at the UE, satisfying $M_{\rm BS} < N_{\rm BS}$ and $M_{\rm UE} <N_{\rm UE}$). The XL-CRIS is comprised of $N_{\rm R}$ reflecting elements arranged on a UCA with radius $r_c$ and half-wavelength inter-element spacing. The azimuth angle of the $n$-th element is represented by $\zeta_n=2\pi(n-1)/N_{\rm R}$. Moreover, for the considered NOMA system, the bandwidth is $f_s$ and the number of subcarriers is $N$. The first $K$ of these subcarriers are allocated as pilots to facilitate channel estimation, where the frequency of the first subcarrier is $f_c$.
\subsection{Channel Modeling}
Owing to the rotational symmetry of XL-CRIS, the boundary of the effective Rayleigh distance is angle-invariant, which can cover a region of a cellular cell and eliminate complex far-field and near-field hybrid channel modeling \cite{10243590}. 
Thus, all UEs in the considering model can be within the near-field region. Besides, the wideband signals will cause beam squint effect, which requires separate channel modeling for different subcarriers \cite{10271123}. Considering that the RIS has a much larger aperture compared to the BS and UE, we adopt a one-sided near-field channel model for the BS-RIS and RIS–UE channels, under which the BS and UE are approximated as point sources. For the BS-RIS channel, due to the elevated deployment of the BS and RIS, the BS–RIS channel can be approximated as quasi-static, where the LoS path is the dominant propagation component. Accordingly, the BS-RIS channel associated with the $k$-th subcarrier can be formulated using the Saleh-Valenzuela model as  
\begin{equation}
{\bf Z}[k] = \alpha e^{-j2 \pi f_k \tau_{\rm BR}} {\bf a}_{\rm R}(\theta_{\rm BR},f_k,d_{\rm BR}) {\bf a}_{\rm B}^T(\phi_{\rm BR},f_k)
\label{G_k.eqn}
\end{equation} 
where $\alpha$, $f_k$, $\theta_{\rm BR}$, $\phi_{\rm BR}$, $\tau_{\rm BR}$, and $d_{\rm BR}$ denote the complex channel gain, the frequency corresponding to the $k$-th subcarrier, the AoA of RIS, AoD of BS, the propagation time delay, and the BS–RIS distance, respectively. Here, $f_k = f_c + \frac{f_s}{N}(k-1), k \in [1,N]$. Here, the array response vectors corresponding to the RIS and BS are denoted as 
\begin{equation}
\begin{aligned}
{\bf a}&_{\rm R}(\theta_{\rm BR},f_k,d_{\rm BR}) = \\
&\frac{1}{\sqrt{N_{\rm R}}}[e^{-j \frac{2 \pi f_k}{v_c}(d_{\rm BR}^{(1)} -d_{\rm BR})}, \cdots, e^{-j \frac{2 \pi f_k}{v_c}(d_{\rm BR}^{(N_{\rm R})} -d_{\rm BR})}]^T
\label{ar_theta.eqn}
\end{aligned}
\end{equation} 
\begin{equation}
\begin{aligned}
{\bf a}&_{\rm B}(\phi_{\rm BR},f_k) = \\ &\frac{1}{\sqrt{N_{\rm BS}}}[1,e^{j\pi \frac{f_k}{f_c} {\rm cos}(\phi_{\rm BR})}, \cdots, e^{j \pi \frac{f_k}{f_c} (N_{\rm BS}-1){\rm cos}(\phi_{\rm BR})}]^T
\label{ab_phi.eqn}
\end{aligned}
\end{equation} 
where $v_c$ denotes the velocity of light and $d_{\rm BR}^{(n)}$ denotes the propagation distance between the BS and the $n$-th RIS element, which is given by
\begin{equation}
\begin{aligned}
d_{\rm BR}^{(n)} = \sqrt{(d_{\rm BR})^2 + (r_c)^2- 2 r_c d_{\rm BR} {\rm cos}(\phi_{\rm BR} - \zeta_n)}
\label{r_n_eqn.eqn}
\end{aligned}
\end{equation} 

By leveraging the second-order Taylor series expansion $\sqrt{1+a} = 1 + \frac{a}{2} - \frac{a^2}{8} + {\cal{O}} (a^3)$, we can expand formula \eqref{r_n_eqn.eqn} as
\begin{equation}
\begin{aligned}
d_{\rm BR}^{(n)} = d_{\rm BR} - r_c {\rm cos} (\theta_{\rm BR} - \zeta_n) + \frac{r_c^2}{2 d_{\rm BR}} {\rm sin}^2(\theta_{\rm BR} - \zeta_n)
\label{r_n_ex}
\end{aligned}
\end{equation} 
Notice that with fixed and known positions for both the BS and RIS, the parameters $\theta_{\rm BR}$, $\phi_{\rm BR}$, $\tau_{\rm BR}$, and $d_{\rm BR}$ are known in the system. 
 
For the RIS-UE channel, we assume there exist multiple individual scattering paths. Therefore, the multi-path geometric mmWave channel model is employed to capture the inherent sparsity of the mmWave MIMO channel. Accordingly, the RIS-UE channel on the $k$-th subcarrier is given by
\begin{equation}
\begin{aligned}
{\bf H}[k] =  \sum_{l=1}^{L} \beta_l e^{-j2 \pi  f_k \tau_{\rm RM}^l} {\bf a}_{\rm U}(\theta_{\rm RM}^l,f_k) {\bf a}_{\rm R}^T(\phi_{\rm RM}^l,f_k,d_{\rm RM}^l)
\label{H_k}
\end{aligned}
\end{equation} 
where $\beta_l$, $\tau_{\rm RM}^l$, $\theta_{\rm RM}^l$, and $\phi_{\rm RM}^l$ denote the complex channel gain, the propagation time delay associated with the $l$-th path of the RIS-UE channel, AoA of UE, and AoD of RIS, respectively. Similar to \eqref{ar_theta.eqn}, \eqref{ab_phi.eqn}, and \eqref{r_n_ex}, the array response vectors of the UE and RIS are given by 
\begin{equation}
\begin{aligned}
{\bf a}&_{\rm U}(\theta_{\rm RM}, f_k) = \\ &\frac{1}{\sqrt{N_{\rm UE}}}[1,e^{j\pi \frac{f_k}{f_c} {\rm cos}(\theta_{\rm RM})}, \cdots, e^{j \pi \frac{f_k}{f_c} (N_{\rm UE}-1){\rm cos}(\theta_{\rm RM})}]^T
\end{aligned}
\label{am_theta.eqn}
\end{equation} 
\begin{equation}
\begin{aligned}
{\bf a}_{\rm R}(\phi_{\rm RM}&,f_k, d_{\rm RM}) = \\
\frac{1}{\sqrt{N_{\rm R}}}[&e^{j \frac{2 \pi f_k}{v_c} r_c  ({\rm cos}(\phi_{\rm RM} - \zeta_1)- \frac{r_c}{2d_{\rm RM}}{\rm sin}^2(\phi_{\rm RM} - \zeta_1))}, \cdots, \\
&e^{j \frac{2 \pi f_k}{v_c} r_c  ({\rm cos}(\phi_{\rm RM} - \zeta_{N_{\rm R}})- \frac{r_c}{2d_{\rm RM}}{\rm sin}^2(\phi_{\rm RM} - \zeta_{N_{\rm R}}))} ]^T
\end{aligned}
\label{ar_phi.eqn}
\end{equation} 

By defining cascade channel as ${\bf O}[k] = {\bf Z}[k]^T \odot {\bf H}[k]$, we have
\begin{equation}
\begin{aligned}
{\bf O}[k] = &\sum_{l=1}^{L} \alpha \beta_l e^{-j2 \pi  (f_c + \frac{f_s}{N}(k-1)) (\tau_{\rm RM}^l + \tau_{\rm BR})}  ({\bf a}_{\rm B}(\phi_{\rm BR},f_k) \otimes \\ 
~~{\bf a}_{\rm U}&(\theta_{\rm RM}^l,f_k) )  ({\bf a}_{\rm R}^T(\theta_{\rm BR}, f_k,d_{\rm BR}) \odot {\bf a}_{\rm R}^T(\phi_{\rm RM}^l, f_k, d_{\rm RM}^l))  \\
 = \sum_{l=1}^{L}  &\rho_l  e^{-j \frac{2 \pi  f_s}{N}(k-1) \tau_l}  {\bf a}_{s}(\phi_{\rm BR},\theta_{\rm RM}^l,f_k) \\
&{\bf a}_{r}^T(\theta_{\rm BR},\phi_{\rm RM}^l,f_k,d_{\rm BR},d_{\rm RM}^l)
\end{aligned}
\label{cascade_channel.eqn}
\end{equation} 
where $\rho_l = \alpha \beta_l e^{-j2 \pi  f_c \tau_l} $, $\tau_l = \tau_{\rm RM}^l + \tau_{\rm BR}$, ${\bf a}_{s}(\phi_{\rm BR},\theta_{\rm RM}^l,f_k) = {\bf a}_{\rm B}(\phi_{\rm BR},f_k) \otimes {\bf a}_{\rm U}(\theta_{\rm RM}^l,f_k) $, and ${\bf a}_{r}(\theta_{\rm BR},\phi_{\rm RM}^l,f_k,d_{\rm BR},d_{\rm RM}^l) = ({\bf a}_{\rm R}^T(\theta_{\rm BR}, f_k,d_{\rm BR}) \odot {\bf a}_{\rm R}^T(\phi_{\rm RM}^l, f_k, d_{\rm RM}^l))^T $ which can be rewritten as
\begin{equation}
\begin{aligned}
&{\bf a}_{r}(\theta_{\rm BR},\phi_{\rm RM}^l,f_k,d_{\rm BR},d_{\rm RM}^l) = \\
&\frac{1}{{N_{\rm R}}}[e^{j \frac{2 \pi f_k}{v_c} r_c  ({\rm cos}(\theta_{\rm BR} - \zeta_1) - \frac{r_c}{2d_{\rm BR}}{\rm sin}^2(\theta_{\rm BR} - \zeta_{1}) + {\rm cos}(\phi_{\rm RM}^l - \zeta_1)} \\
& ^{- \frac{r_c}{2d_{\rm RM}^l}{\rm sin}^2(\phi_{\rm RM}^l - \zeta_1))}, \cdots, e^{j  \frac{2 \pi f_k}{v_c} r_c  ({\rm cos}(\theta_{\rm BR} - \zeta_{N_{\rm R}})- \frac{r_c}{2d_{\rm BR}}}\\
& ^{{\rm sin}^2(\theta_{\rm BR} - \zeta_{N_{\rm R}}) + {\rm cos}(\phi_{\rm RM}^l - \zeta_{N_{\rm R}})- \frac{r_c}{2d_{\rm RM}^l}{\rm sin}^2(\phi_{\rm RM}^l - \zeta_{N_{\rm R}}))}]^T
\end{aligned}
\label{ar_rewritten.eqn}
\end{equation} 

\subsection{Signal Modeling}
Within the considered NOMA framework, the BS serves multiple UEs simultaneously using the same time-frequency resource block. Furthermore, quadrature amplitude modulation (QAM) technology is applied to achieve a higher transmission rate. Assuming there are $m$ time slots used for channel estimation, each of which contains $N_b$ NOMA symbols. Then, for the $i$-th UE, the modulated data symbols transmitted via the $n_s$-th parallel data stream on subcarrier $k$ during the $nb$-th NOMA symbol is represented as $x_{i,n_s,nb,m,k}$. Subsequently, the superposed signal $x_{n_s,nb,m,k}$ can be obtained by
\begin{equation}
x_{n_s,nb,m,k} = \sum_{i=1}^{N_{\rm U}} \sqrt{\iota_i P_t}x_{i,n_s,nb,m,k} 
\label{superposed_signal.eqn}
\end{equation} 
where $N_{\rm U}$ represents the total number of UEs sharing a same time-frequency resource block, $P_t$ denotes the total power, and $\iota_i$ indicates the power fraction for the $i$-th UE. Consequently, the signal received at the $k$-th subcarrier during the $nb$-th NOMA symbol of the $m$-th time slot, denoted as ${\bf y}_{nb,m,k}$, is given by
\begin{equation}
\begin{aligned}
{\bf y}_{nb,m,k} =& {\bf W}^T {\bf H}[k] {\rm diag}({\boldsymbol{\Psi}_{nst}}) {\bf G}_k {\bf F} {\bf x}_{nb,m,k} \\
&+{\bf W}^T{\bf n}_{nb,m,k} \in \mathbb{C}^{N_s \times 1}
\end{aligned}
\label{recieved_signal.eqn}
\end{equation} 
where ${\bf F} = {\bf F}_{\rm RF} {\bf F}_{\rm BB} \in {\mathbb{C}}^{N_{\rm BS} \times N_{s}}$ corresponds to the hybrid precoding matrix at the BS, formed by the analog RF precoder ${\bf F}_{\rm RF}$ and the digital baseband precoder ${\bf F}_{\rm BB}$, ${\bf W} = {\bf W}_{\rm RF} {\bf W}_{\rm BB} \in {\mathbb{C}}^{N_{\rm MS} \times N_{s}}$ similarly refers to the hybrid combining matrix at the UE, which is composed of the analog combiner ${\bf W}_{\rm RF}$ and the digital combiner  ${\bf W}_{\rm BB}$. ${\bf x}_{nb,m,k} = [x_{1,nb,m,k},\cdots,x_{N_s,nb,m,k}]^T \in \mathbb{C}^{N_s \times 1}$ represents the parallel data streams transmitted on the $k$-th subcarrier of the $nb$-th NOMA symbol during the $m$-th time slot, $\boldsymbol{\Psi}_m = [e^{j \xi_1},e^{j \xi_2},\cdots,e^{j \xi_{N_{\rm R}}}]^T$, with $\xi_i \in [0,2\pi]$, denotes the RIS phase shift vector at the $m$-th time slot, and ${\bf n}_{nb,m,k} \in \mathbb{C}^{N_{\rm MS} \times 1} \sim {\cal CN}(0,\sigma^2 {\bf I}_{N_{\rm MS}})$ is the additive independent and identically distribution (i.i.d.) zero-mean circularly symmetric complex white Gaussian noise vector, obeying variance $\sigma^2$. It is assumed that $\bf W$ and $\bf F$ stay constant for $m$ time slots.
 
Vectorizing \eqref{recieved_signal.eqn}, we can obtain 
\begin{equation}
\begin{aligned}
&{\bf y}_{nb,m,k}  =  {\rm vec}({\bf y}_{nb,m,k}) = (({\bf F} {\bf x}_{nb,m,k}) \otimes {\bf W})^T\\
&~~~~ ({\bf Z}[k]^T \odot {\bf H}[k]) \boldsymbol{\Psi}_{m} +{\bf W}^T{\bf n}_{nb,m,k} \in \mathbb{C}^{N_s \times 1}
\end{aligned}
\label{vec_recieved_signal.eqn}
\end{equation} 

Let ${\bf X}_{m,k} = [{\bf x}_{1,m,k},\cdots,{\bf x}_{N_b,m,k}] \in \mathbb{C}^{N_s \times N_b}$ denote the matrix stacked by the transmitted NOMA symbols at the pilot subcarriers within a time slot. Then, we can obtain
\begin{equation}
\begin{aligned}
{\bf y}_{m,k} & =  (({\bf F} {\bf X}_{m,k}) \otimes {\bf W})^T({\bf Z}[k]^T \odot {\bf H}[k]) \boldsymbol{\Psi}_{m}\\
&~~+ {\rm vec}({\bf W}^T{\bf N}_{m,k} ) \in \mathbb{C}^{ N_s N_b \times 1},  k = 1,\cdots,K \\
&= {\bf g}_{m,k} + {\rm vec}({\bf W}^T{\bf N}_{m,k} )
\end{aligned}
\label{vec_recieved_signal_collecting.eqn}
\end{equation} 
where ${\bf y}_{m,k} = [{\bf y}^T_{1,m,k},\cdots,{\bf y}^T_{N_b,m,k}]^T$ denotes the stacked received signal, ${\bf N}_{m,k} = [{\bf n}_{1,m,k},\cdots, {\bf n}_{N_b,m,k}] \in \mathbb{C}^{N_{\rm MS} \times N_b}$ denotes noise matrices, and ${\bf g}_{m,k}$ denotes the noiseless received signal. Across different time slots, the $N_s$ parallel streams on the pilot subcarriers remain unchanged, though they are independent within the same time slot. Hence, we let ${\bf X}  = {\bf X}_{m,k}$ and define $\boldsymbol{\Upsilon} = ({\bf FX}) \otimes {\bf W}$. Upon stacking the RIS phase-shift vectors for a time slot as $\boldsymbol{\Xi} = [\boldsymbol{\Psi}_{1},\cdots,\boldsymbol{\Psi}_{M}]$, the received signal at the $k$-th pilot subcarrier is given by
\begin{equation}
\begin{aligned}
{\bf Y}_{k} & =  \boldsymbol{\Upsilon}^T({\bf Z}[k]^T \odot {\bf H}[k]) \boldsymbol{\Xi}\\
&~~~~ +{\rm Vec}^1_2(\boldsymbol{{\cal N}}_{k} \times_1 {\bf W}^T) \in \mathbb{C}^{N_s N_b \times M}
\end{aligned}
\label{matrix_recieved_signal_collecting.eqn}
\end{equation} 
where ${\bf Y}_{k} = [{\bf y}_{1,m,k}, \cdots, {\bf y}_{M,k}]$ denotes the stacked received signal matrix, ${\rm Vec}^1_2(\cdot)$ denotes the process of stacking the vectorized frontal slices of a tensor. As an example, consider a third-order tensor $\boldsymbol{{\cal A}} \in \mathbb{C}^{I_1 \times I_2 \times I_3}$, ${\rm Vec}^1_2(\boldsymbol{{\cal A}}) = [{\rm vec}(\boldsymbol{{\cal A}}(:,:,1)),\cdots, [{\rm vec}(\boldsymbol{{\cal A}}(:,:,I_3))] \in \mathbb{C}^{I_1 I_2 \times I_3}$, and $\boldsymbol{{\cal N}}_{k} \in \mathbb{C}^{N_{\rm MS} \times N_b \times M}$. 

Substituting \eqref{cascade_channel.eqn} into \eqref{matrix_recieved_signal_collecting.eqn}, the received signal on the $k$-th pilot subcarrier can be represented as 
\begin{equation}
\begin{aligned}
{\bf Y}_{k} &=  \sum_{l=1}^{L} \rho_l e^{-j2 \pi  f_k \tau_l} \boldsymbol{\Upsilon}^T ({\bf a}_{s}(\phi_{\rm BR},\theta_{\rm RM}^l,f_k) \\
&~~~~{\bf a}_{r}^T(\theta_{\rm BR},\phi_{\rm RM}^l,f_k,d_{\rm BR},d_{\rm RM}^l)) \boldsymbol{\Xi} + {\rm Vec}^1_2(\boldsymbol{{\cal N}}_{k} \times_1 {\bf W}^T) \\
&= {\bf A}_{s,k} {\bf \Sigma}_k {\bf A}_{r,k} ^T + {\rm Vec}^1_2(\boldsymbol{{\cal N}}_{k} \times_1 {\bf W}^T)
\end{aligned}
\label{recieved_signal_matrix_rewrite.eqn}
\end{equation} 
where 
\begin{equation}
\begin{aligned}
{\bf A}_{s,k} &= [\boldsymbol{\Upsilon}^T {\bf a}_{s}(\phi_{\rm BR},\theta_{\rm RM}^1,f_k),\cdots,\boldsymbol{\Upsilon}^T {\bf a}_{s}(\phi_{\rm BR},\theta_{\rm RM}^L,f_k)] \\
{\bf \Sigma}_k & = {\rm diag}(\rho_1 e^{-j2 \pi  f_k \tau_1},\cdots,\rho_L e^{-j2 \pi  f_k \tau_L}) \\
{\bf A}_{r,k} & = [\boldsymbol{\Xi}^T {\bf a}_{r}(\theta_{\rm BR},\phi_{\rm RM}^1,f_k,d_{\rm BR},d_{\rm RM}^1)),\cdots, \\
&~~~~~\boldsymbol{\Xi}^T {\bf a}_{r}(\theta_{\rm BR},\phi_{\rm RM}^L,f_k,d_{\rm BR},d_{\rm RM}^L))]
\end{aligned}
\label{ASK_ARK}
\end{equation} 

Subsequently, stacking the received signals for the first $K$ subcarriers gives
\begin{equation}
\begin{aligned}
\boldsymbol{{\cal Y}} & = \boldsymbol{{\cal G}} + \boldsymbol{{\cal N}}^{\bf W}
\end{aligned}
\label{tensor_recieved_signal.eqn}
\end{equation} 
where $\boldsymbol{{\cal Y}}(:,:,k) = {\bf Y}_k$, $\boldsymbol{{\cal G}}(:,:,k) = {\bf A}_{s,k} {\bf \Sigma}_k {\bf A}_{r,k} ^T$, and $\boldsymbol{{\cal N}}^{\bf W} = {\rm Vec}^1_2(\boldsymbol{{\cal N}} \times_1 {\bf W}^T) $ with $\boldsymbol{{\cal N}}(:,:,:,k) = \boldsymbol{{\cal N}}_{k} $. Notably, if the beam squint effect is ignored, the received signal $\boldsymbol{{\cal Y}} $ can be represented as a low-rank tensor that satisfies CP decomposition form. However, in wideband system, the low-rank property is destroyed, and some state-of-the-art methods based on low-rank tensor decomposition will no longer be applicable.
\section{Near-Field Channel Estimation}
In this section, we propose a multi-stage near-field channel estimation (MSNFCE) method to restore the geometric channel parameters. In the first stage, the MUSIC method with multi-subcarrier observation aggregation is exploited to estimate $\boldsymbol{\theta}_{\rm RM}$. Then, a Nelder-Mead simplex refinement-assisted correlation-based estimator is applied to estimate $\boldsymbol{\phi}_{\rm RM}$ and $\boldsymbol{d}_{\rm RM}$ in the second stage. In the third stage, by leveraging the Vandermonde structure of the constructed matrix, ${\boldsymbol{\tau }}_l$ can be estimated. Finally, we construct a least squares problem to estimate the cascade channel gain based on the estimated values above.

\subsection{Estimation of $\boldsymbol{\theta}_{\rm RM}$}
According to \eqref{recieved_signal_matrix_rewrite.eqn} and \eqref{ASK_ARK}, for each subcarrier, ${\bf A}_{s,k}$ is a beam space manifold matrix with $\boldsymbol{\Upsilon}^T {\bf a}_{s}(\phi_{\rm BR},\theta_{\rm RM}^l,f_k)$ as its column, while ${\bf \Sigma}_k {\bf A}_{r,k} ^T$ can be regarded as the equivalent received signal. Accordingly, the covariance matrix for the received signal on each subcarrier is given by
\begin{equation}
\begin{aligned}
{\bf R}_k &= \mathbb{E}[{\bf Y}_{k} {\bf Y}_{k}^H] \\
&={\bf A}_{s,k} \mathbb{E}[ {\bf \Sigma}_k {\bf A}_{r,k}^T   {\bf A}_{r,k}^*  {\bf \Sigma}_k^T ] {\bf A}_{s,k}^H + \sigma^2 {\bf I}_{N_s N_b}
\end{aligned}
\label{covariance_matrix}
\end{equation} 

Subsequently, performing the eigenvalue decomposition (EVD) on matrix ${\bf R}_k = {\bf Q}_k {\bf \Lambda}_k {\bf Q}_k^H$ with the eigenvalues sorted in descending order yields bases for the beamspace signal and noise subspaces, where ${\bf Q}_k = [{\bf q}_{1,k},{\bf q}_{2,k},\cdots,{\bf q}_{N_s N_b,k}]$. Let ${\bf Q}_{s,k}$ and ${\bf Q}_{n,k}$ be orthogonal matrix associated with the $k$-th subcarrier, which span the beamspace signal and noise subspaces, respectively, denoted as
\begin{equation}
\begin{aligned}
{\bf Q}_{s,k} &= [{\bf q}_{1,k},{\bf q}_{2,k},\cdots, {\bf q}_{L,k}] \\
{\bf Q}_{n,k} &= [{\bf q}_{L+1,k},{\bf q}_{L+2,k},\cdots, {\bf q}_{N_s N_b,k}]
\end{aligned}
\label{orthogonal_basis}
\end{equation}

Accordingly, the MUSIC spectrum associated with the $k$-th subcarrier is given by 
\begin{equation}
\begin{aligned}
&S_k(\theta_{\rm RM}^{l,k}) = \\
&\frac{1}{{\bf a}_{s}^H(\phi_{\rm BR},\theta_{\rm RM}^{l,k},f_k) \boldsymbol{\Upsilon}^* {\bf Q}_{n,k}{\bf Q}_{n,k}^H \boldsymbol{\Upsilon}^T {\bf a}_{s}(\phi_{\rm BR},\theta_{\rm RM}^{l,k},f_k)}
\end{aligned}
\label{MUSIC_specrum}
\end{equation}

Subsequently, by applying a one-dimensional search on \eqref{MUSIC_specrum} with $L$ peak vales, we can obtain $\hat{\boldsymbol{\theta}}_{\rm RM}^k = \{\hat{\theta}_{\rm RM}^{l,k} \}_{l=1}^L$. To achieve more accurate and stable estimation, the $\hat{\theta}_{\rm RM}^{l,k}$ is estimated separately for each of the $K$ pilot subcarriers. However, the angle estimates from each subcarrier suffer from a permutation ambiguity and may exist estimation anomalies problem at low signal-to-noise ratio (SNR) scenarios. Therefore, we apply a classical K-means clustering algorithm to aggregate the angle estimates into $L$ classes firstly. Then, for each class, we use one-Sigma standard deviation criterion to remove the outliers. Finally, by averaging  the processed data in each class, we can obtain the final estimated AoA of UE $\tilde{\theta}_{\rm RM}^{l}$. Here, we prior knowledge of the number of paths $L$. If the rank of ${\bf Y}_{k}$, i.e. the number of paths, is unknown, we can adopt the singular value decomposition and set a threshold to determine the rank of ${\bf Y}_{k}$ of each pilot subcarrier, then extract the rank with the maximum occurrence frequency.

\subsection{Estimation of $\boldsymbol{\phi}_{\rm RM}$ and $\boldsymbol{d}_{\rm RM}$}
Based on the estimated $\tilde{\theta}_{\rm RM}^{l}$, we can reconstruct $\tilde{\bf A}_{s,k}$, i.e. $\tilde{\bf A}_{s,k} = [\boldsymbol{\Upsilon}^T {\bf a}_{s}(\phi_{\rm BR},\tilde{\theta}_{\rm RM}^1,f_k),\cdots,\boldsymbol{\Upsilon}^T {\bf a}_{s}(\phi_{\rm BR},\tilde{\theta}_{\rm RM}^L,f_k)]$ whose relationship with the true factor matrix can be expressed as
\begin{equation}
\begin{aligned}
\tilde{\bf A}_{s,k} = {\bf A}_{s,k} {\bf \Omega}
\end{aligned}
\label{factor_relationship}
\end{equation}
where ${\bf \Omega}$ is the unknown permutation matrix. Here, to simplify the derivation process, we neglect the influence of noise. Then, according to \eqref{recieved_signal_matrix_rewrite.eqn} and \eqref{factor_relationship}, we can derive 
\begin{equation}
\begin{aligned}
{\bf P}_k&= [{\bf p}_{k,1},{\bf p}_{k,2},\cdots,{\bf p}_{k,L}] =
(\tilde{\bf A}_{s,k}^{\dagger}  {\bf Y}_{k} )^T \\
& = 
{\bf A}_{r,k}    {\bf \Sigma}_k {\bf \Omega}^{-T}  = ({\bf A}_{r,k}   {\bf \Sigma}_k)  {\bf \Omega} \\
&  = \tilde{\bf A}_{r,k} \tilde{\bf \Sigma}_{k}
\end{aligned}
\label{A_rk_est}
\end{equation}
where $\tilde{\bf A}_{s,k}$ and $\tilde{\bf A}_{r,k}$ share the same permutation matrix, and $ \tilde{\bf \Sigma}_{k}$ represents the diagonal matrix whose diagonal elements are permuted in the same order as  $ {\bf \Omega}$. It is worth mentioning that by using the same permutation matrix, the geometric channel parameters corresponding to each path are automatically paired, avoiding the matching problem.

For each column of ${\bf P}_k$, we can regard it as the permuted column of ${\bf A}_{r,k}$ multiplying a scaling coefficient. Accordingly, by developing a correlation-based estimator, parameters ${\tilde{\phi}}_{\rm RM}^{l,k}$ and $\tilde{d}_{\rm RM}^{l,k}$ can be estimated by
\begin{equation}
\begin{aligned}
{\hat{\phi}}&_{\rm RM}^{l,k} ,\hat{d}_{\rm RM}^{l,k} = \\
&\underset{\phi_{\rm RM}^l,d_{\rm RM}^l}{\rm argmax} \; \frac{\vert {\bf p}_{k,l}^H \boldsymbol{\Xi}^T {\bf a}_{r}(\theta_{\rm BR},\phi_{\rm RM}^l,f_k,d_{\rm BR},d_{\rm RM}^l) \vert}{\Vert {\bf p}_{k,l} \Vert_2 \Vert \boldsymbol{\Xi}^T {\bf a}_{r}(\theta_{\rm BR},\phi_{\rm RM}^l,f_k,d_{\rm BR},d_{\rm RM}^l \Vert_2}
\end{aligned}
\label{AOD_DIS}
\end{equation}

Obviously, problem \eqref{AOD_DIS} can be accurately resolved via a two-dimensional search over a finely discretized grid. However, the large scale grids will poses a significant computational challenge. Thus, a Nelder-Mead simplex refinement-assisted two-dimensional search method is applied to maintain the estimation accuracy while reducing the computational complexity. In specific, an initial coarse two-dimensional search over a coarse grid is performed to obtain ${\hat{\phi}}_{\rm RM}^{l,k} ,\hat{d}_{\rm RM}^{l,k}$. Notice that using only a single pilot subcarrier for estimation may result in unstable initial values, potentially leading to the refinement algorithm stagnating around a local optimum or exhibiting slow convergence. Hence, we use all or part of the pilot subcarriers for estimation, and then use one-Sigma standard deviation criterion to remove the outliers. Subsequently, we average the processed estimates to obtain the final initial solution ${\hat{\phi}}_{\rm RM}^{l}$ and $\hat{d}_{\rm RM}^{l}$. Then, a nonlinear optimization problem is constructed for solution refinement, which is expressed as

\noindent
\begin{equation}
\begin{aligned}
&{\tilde{\phi}}_{\rm RM}^{l} ,\tilde{d}_{\rm RM}^{l} = \\
&\underset{\phi_{\rm RM}^l,d_{\rm RM}^l}{\rm argmin} \; \sum_{k=1}^{K}  \frac{- \vert {\bf p}_{k,l}^H \boldsymbol{\Xi}^T {\bf a}_{r}(\theta_{\rm BR},\phi_{\rm RM}^l,f_k,d_{\rm BR},d_{\rm RM}^l) \vert}{\Vert {\bf p}_{k,l} \Vert_2 \Vert \boldsymbol{\Xi}^T {\bf a}_{r}(\theta_{\rm BR},\phi_{\rm RM}^l,f_k,d_{\rm BR},d_{\rm RM}^l \Vert_2}
\end{aligned}
\label{AOD_DIS_refinement}
\end{equation}
Problem \eqref{AOD_DIS_refinement} can be efficiently solved through a Nelder-Mead simplex method \cite{nelder1965simplex} using the obtained initial solution ${\hat{\phi}}_{\rm RM}^{l}$ and $\hat{d}_{\rm RM}^{l}$. Owing to its derivative-free iterative structure, the Nelder–Mead simplex method is capable of achieving fast convergence in nonlinear optimization problems \cite{lagarias1998convergence}. Accordingly, we can obtain the exact solution while significantly reducing computational complexity.
\subsection{Estimation of $\boldsymbol{\tau}$}
Based on the estimated ${\tilde{\phi}}_{\rm RM}^{l}$ and $\tilde{d}_{\rm RM}^{l}$, we can reconstruct $\tilde{\bf A}_{r,k}$, i.e. $\tilde{\bf A}_{r,k} = [\boldsymbol{\Xi}^T {\bf a}_{r}(\theta_{\rm BR},\tilde{\phi}_{\rm RM}^1,f_k,d_{\rm BR},\tilde{d}_{\rm RM}^1),\cdots,$ $\boldsymbol{\Xi}^T {\bf a}_{r}(\theta_{\rm BR},\tilde{\phi}_{\rm RM}^L,f_k,d_{\rm BR},\tilde{d}_{\rm RM}^L)]$. Due to automatic pairing, we can derive permuted matrix $\tilde{\bf \Sigma}_{k}$ based on \eqref{A_rk_est}, i.e.
\begin{equation}
\begin{aligned}
\tilde{\bf \Sigma}_{k} = \tilde{\bf A}_{r,k}^{\dagger} {\bf P}_k = {\rm diag}(\tilde{\rho}_1 e^{-j2 \pi  f_k \tilde{\tau}_1},\cdots,\tilde{\rho}_L e^{-j2 \pi  f_k \tilde{\tau}_L})
\end{aligned}
\label{tau_cascade_gain}
\end{equation}

By collecting the $l$-th diagonal element of $\tilde{\bf \Sigma}_{k}$ under all the pilot subcarriers, we have  
\begin{equation}
\begin{aligned}
{\bf s}_l = [\tilde{\rho}_l e^{-j2 \pi  f_1 \tilde{\tau}_l},\cdots,\tilde{\rho}_l e^{-j2 \pi  f_K \tilde{\tau}_l} ]^T \in \mathbb{C}^{K \times 1}
\end{aligned}
\label{tau_cascade_collect}
\end{equation}

Accordingly, we can obtain ${\bf S} = [{\bf s}_1,\cdots,{\bf s}_L]$ which has Vandermonde structure. Then, by performing element-wise division, i.e. $s_l(i+1) / s_l(i)$, we can obtain
\begin{equation}
\begin{aligned}
\hat{\bf s}_l = [e^{-j2 \pi  \frac{f_s}{N} \tilde{\tau}_l},\cdots, e^{-j2 \pi  \frac{f_s}{N} \tilde{\tau}_l} ]^T \in \mathbb{C}^{(K-1) \times 1}
\end{aligned}
\label{tau_collect}
\end{equation}

Subsequently, averaging all elements in $\hat{\bf s}_l$ and dividing $e^{-j2 \pi f_s {\tau}_{\rm BR} / N}$, we have $t_l = e^{-j2 \pi f_s {\tau}_{\rm RU} / N}$. Accordingly, $\tilde{\tau}_l$ can be directly estimated by
\begin{equation}
\begin{aligned}
\tilde{\tau}_l = \tau_{\rm BR} + \frac{1}{2 \pi f_s N}  \angle (t_l )
\end{aligned}
\label{tau_est}
\end{equation}
where $\angle(\cdot)$ denotes the phase extraction operation.

\subsection{Estimation of $\boldsymbol{\rho}$}
Using the previously estimated channel parameters, we construct a least-squares problem to estimate the permuted cascade channel gain $\boldsymbol{\tilde{\rho}}_{k} =[\tilde{\rho}_{1,k},\cdots,\tilde{\rho}_{L,k}]^T$ at the $k$-th pilot subcarrier, expressed as
\begin{equation}
\begin{aligned}
\tilde{\rho}_{l,k} =\underset{\rho_l}{\rm argmin} \; &\Vert {\bf Y}_{k} -  \sum_{l=1}^{L} \rho_l \boldsymbol{\Phi}_{l,k} \Vert_F^2 
\end{aligned}
\label{LS_rewritten.eqn}
\end{equation}
where $\boldsymbol{\Phi}_{l,k} = e^{-j2 \pi \frac{k}{N} f_s \tilde{\tau}_l} \boldsymbol{\Upsilon}^T {\bf a}_{s}(\phi_{\rm BR},\tilde{\theta}_{\rm RM}^l,f_k)$ ${\bf a}_{r}^T(\theta_{\rm BR},\phi_{\rm RM}^l,f_k,d_{\rm BR},d_{\rm RM}^l)) \boldsymbol{\Xi}$. By expanding \eqref{LS_rewritten.eqn}, we can derive
\begin{equation}
\begin{aligned}
&\tilde{\rho}_{l,k} = \underset{\rho_l}{\rm argmin} \; {\rm Tr}(({\bf Y}_k -  \sum_{l=1}^{L} \rho_l \boldsymbol{\Phi}_{l,k})({\bf Y}_k -  \sum_{l=1}^{L} \rho_l \boldsymbol{\Phi}_{l,k})^H) \\
& = \underset{\rho_l}{\rm argmin} \; {\rm Tr}( {\bf Y}_k  {\bf Y}_k^H - \sum_{l=1}^{L} \rho_l^*{\bf Y}_k \boldsymbol{\Phi}_{l,k}^H- \sum_{l=1}^{L} \rho_l \boldsymbol{\Phi}_{l,k} {\bf Y}_k^H)  \\
&\qquad \qquad \; + {\rm Tr}(\sum_{l_1=1}^{L}\sum_{l_2=1}^{L} \rho_{l_1}\rho_{l_2}^* \boldsymbol{\Phi}_{l_1,k} \boldsymbol{\Phi}_{l_2,k}^H) 
\end{aligned}
\label{objective_function.eqn}
\end{equation}
Defining matrix $\boldsymbol{\Gamma}_{k}$ whose entries $\boldsymbol{\Gamma}_{k}^{(i,j)}$ are equal to ${\rm Tr}(\boldsymbol{\Phi}_{j,k} \boldsymbol{\Phi}_{i,k}^H)$, problem \eqref{objective_function.eqn} can be rewritten as a quadratic form
\begin{equation}
\begin{aligned}
\tilde{\boldsymbol{\rho}}_k&= \underset{\boldsymbol{\rho}}{\rm argmin} \; \boldsymbol{\rho}^H
\begin{bmatrix} 
{\rm Tr}(\boldsymbol{\Phi}_{1,k}\boldsymbol{\Phi}_{1,k}^H)  & \cdots & {\rm Tr}(\boldsymbol{\Phi}_{L,k}\boldsymbol{\Phi}_{1,k}^H)\\
\vdots &\ddots & \vdots\\
{\rm Tr}(\boldsymbol{\Phi}_{1,k}\boldsymbol{\Phi}_{L,k}^H)  & \cdots & {\rm Tr}(\boldsymbol{\Phi}_{L,k}\boldsymbol{\Phi}_{L,k}^H)
\end{bmatrix} \boldsymbol{\rho} - \\
&\qquad \qquad \; \boldsymbol{\rho}^H [{\rm Tr}({\bf Y}_k \boldsymbol{\Phi}_{1,k}^H),\cdots,{\rm Tr}({\bf Y}_k \boldsymbol{\Phi}_{L,k}^H)]^T - \\
& \qquad \qquad \;  [{\rm Tr}({\bf Y}_k \boldsymbol{\Phi}_{1,k}^H),\cdots,{\rm Tr}({\bf Y}_k \boldsymbol{\Phi}_{L,k}^H)]^H \boldsymbol{\rho} \\
& = \underset{\boldsymbol{\rho}}{\rm argmin} \; \boldsymbol{\rho}^H \boldsymbol{\Gamma}_k \boldsymbol{\rho} - \boldsymbol{\rho}^H\boldsymbol{\zeta}_k - \boldsymbol{\zeta}_k^H  \boldsymbol{\rho}
\end{aligned}
\label{objective_function_new.eqn}
\end{equation}
where $\boldsymbol{\zeta}_{k} = $ $[{\rm Tr}({\bf Y}_{k} \boldsymbol{\Phi}_{1,k}^H),\cdots,{\rm Tr}({\bf Y}_{k} \boldsymbol{\Phi}_{L,k}^H)]^T$.

Apparently, the solution to problem \eqref{objective_function_new.eqn} can be calculated by
\begin{equation}
\begin{aligned}
\tilde{\boldsymbol{\rho}}_k = \boldsymbol{\Gamma}_k^{-1} \boldsymbol{\zeta}_k
\end{aligned}
\label{objective_function_solution.eqn}
\end{equation}

Then, we average the estimates $\boldsymbol{\tilde{\rho}}_{k}$ for $k = 1,\cdots, K$ to obtain the final permuted cascade channel gain $\boldsymbol{\tilde{\rho}} = [\tilde{\rho}_1,\cdots,\tilde{\rho}_L]^T$. Overall, the overall procedure of the proposed method is summarized in Algorithm \ref{MSNFCEE_algorithm}.
\begin{algorithm}[htbp]
	\caption{MSNFCE Algorithm}
	\label{MSNFCEE_algorithm}
	\begin{algorithmic}  
		\State {\bf Inputs}: 
		Observed noisy signal tensor $\boldsymbol{{\cal Y}}$.
		
		\begin{itemize}
			\item[{\bf 1}:] Estimation of $\boldsymbol{\theta}_{\rm RM}$
			\begin{itemize}
				\item[{\bf 1)}]  Calculate covariance matrix ${\bf R}_k$ by \eqref{covariance_matrix}.
				
				\item[{\bf 2)}] Compute the EVD of ${\bf R}_k$ to obtain the signal and noise subspaces.
				
				\item[{\bf 3)}] Estimate $\theta_{\rm RM}^{l,k}$ by \eqref{MUSIC_specrum}.
				
				\item[{\bf 4)}] Perform clustering and fusion averaging operations to obtain $\tilde{\theta}_{\rm RM}^{l}$.	
			\end{itemize}
			\item[{\bf 2}:] Estimation of $\boldsymbol{\phi}_{\rm RM}$ and $\boldsymbol{d}_{\rm RM}$
			\begin{itemize}
				\item[{\bf 1)}] Reconstruct $\tilde{\bf A}_{s,k}$ and obtain ${\bf P}_k$ by \eqref{A_rk_est}.
				
				\item[{\bf 2)}] Obtain the initial solution ${\hat{\phi}}_{\rm RM}^{l}$ and $\hat{d}_{\rm RM}^{l}$ by outlier removal of the solution to \eqref{AOD_DIS} and fusion averaging. 
				
				\item[{\bf 3)}] Refine the solution to obtain ${\tilde{\phi}}_{\rm RM}^{l}$ and $\tilde{d}_{\rm RM}^{l}$ by \eqref{AOD_DIS_refinement}.
			\end{itemize}
		\item[{\bf 3}:] Estimation of $\boldsymbol{\tau}$
		\begin{itemize}
			\item[{\bf 1)}] Reconstruct $\tilde{\bf A}_{r,k}$ and obtain $\tilde{\bf \Sigma}_{k}$ by \eqref{tau_cascade_gain}.
			
			\item[{\bf 2)}] Estimate $\tilde{\tau}_l$ by \eqref{tau_est}.
		\end{itemize}
		
		\item[{\bf 4}:] Estimation of $\boldsymbol{\rho}$
		\begin{itemize}
			\item[{\bf 1)}] Estimate $\tilde{\boldsymbol{\rho}}_k$  by \eqref{objective_function_solution.eqn}.
			
			\item[{\bf 2)}] Obtain $\tilde{\rho}_l$ by fusion averaging. 
		\end{itemize}
		\end{itemize}
		\State{\bf Output}: Channel parameters $\{ \tilde{\theta}_{\rm RM}^{l}, {\tilde{\phi}}_{\rm RM}^{l}, \tilde{d}_{\rm RM}^{l} \tilde{\tau}_l,  \tilde{\rho}_l \}_{l=1}^{L}$ and the cascade channel $\tilde{{\bf O}}[k]$ obtained by substituting all the estimated channel parameters into \eqref{cascade_channel.eqn}. 
	\end{algorithmic}
\end{algorithm}	

\section{CRB Analysis} \label{CRB_ana}
\begin{figure*}[b]
	\centering
	\subfigure{
		\begin{minipage}[h]{0.33\linewidth}
			\centering
			\includegraphics[width=2.3in]{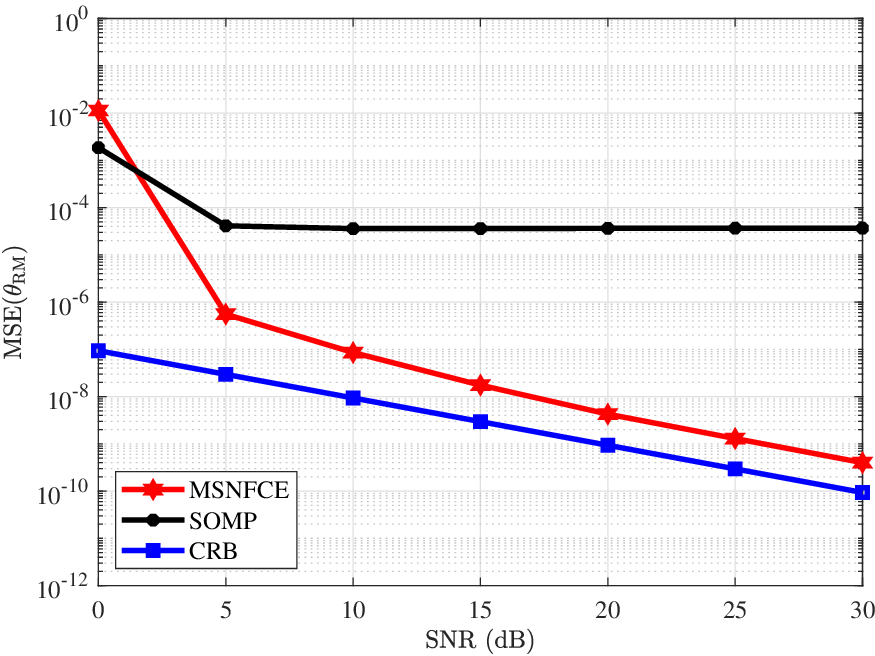}		
			\caption*{(a) $\rm MSE$ of $\boldsymbol{\theta}_{\rm RM}$}
			\label{F1_theta_RM}
		\end{minipage}%
	}%
	\subfigure{
		\begin{minipage}[h]{0.33\linewidth}
			\centering
			\includegraphics[width=2.3in]{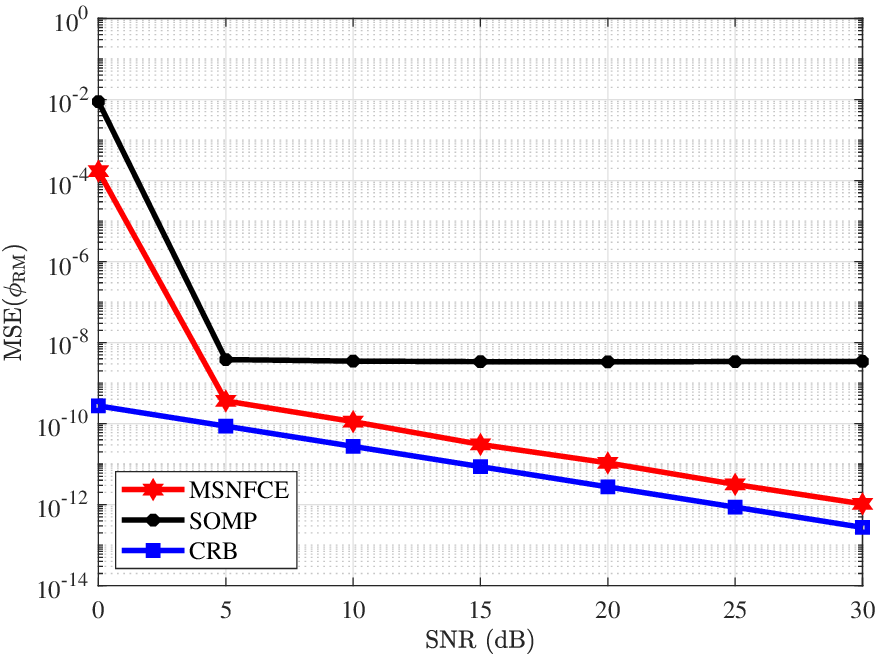}
			\caption*{(b) $\rm MSE$ of $\boldsymbol{\phi}_{\rm RM}$}
			\label{F1_phi_RM}
		\end{minipage}%
	}%
	\subfigure{
		\begin{minipage}[h]{0.33\linewidth}
			\centering
			\includegraphics[width=2.3in]{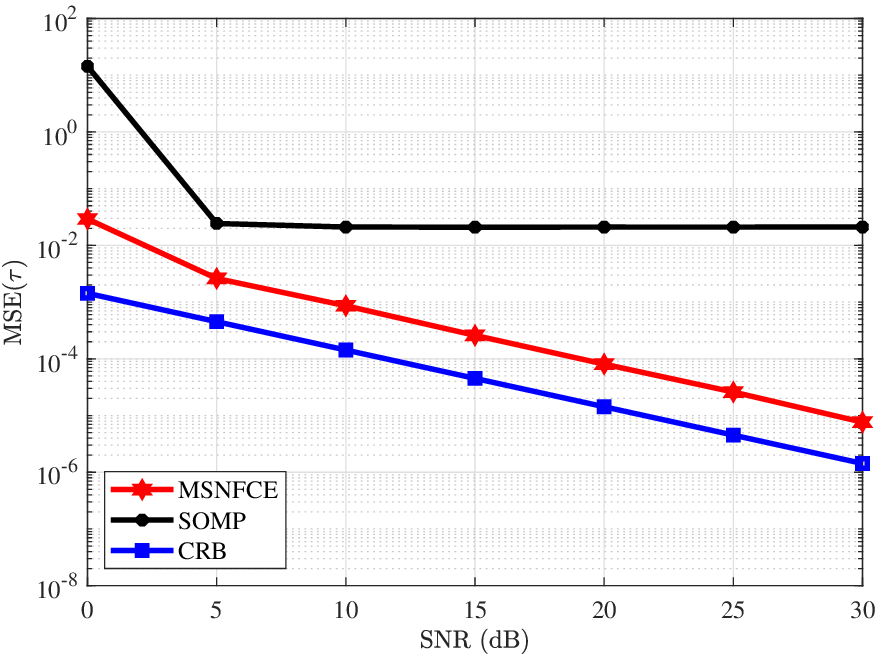}
			\caption*{(c) $\rm MSE$ of $\boldsymbol{\tau}$}
			\label{F1_tau}
		\end{minipage}%
	}%
	
	\subfigure{
		\begin{minipage}[h]{0.33\linewidth}
			\centering
			\includegraphics[width=2.3in]{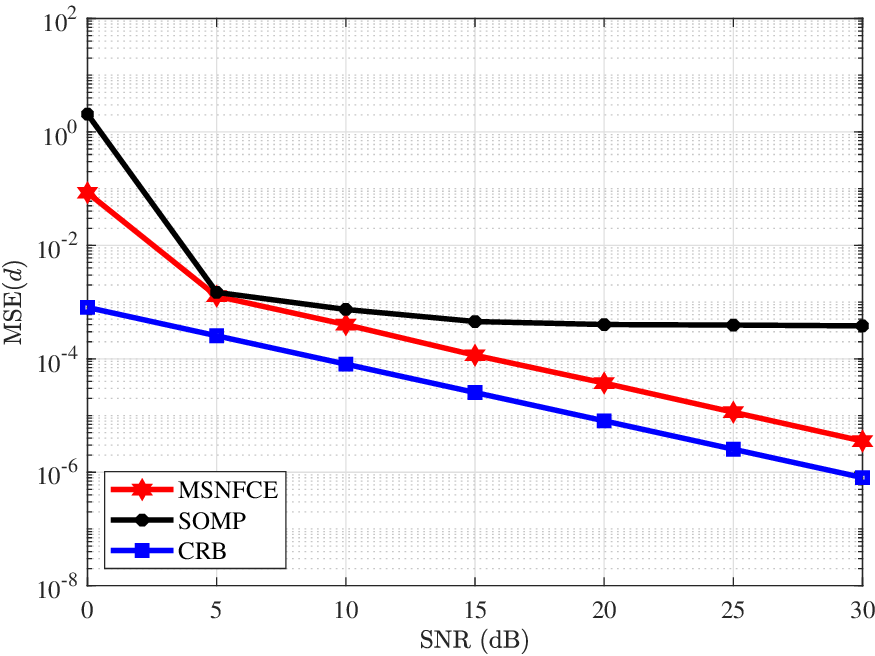}
			\caption*{(d) $\rm MSE$ of $\boldsymbol{d}$}
			\label{F1_d}
		\end{minipage}%
	}%
	\subfigure{
		\begin{minipage}[h]{0.33\linewidth}
			\centering
			\includegraphics[width=2.3in]{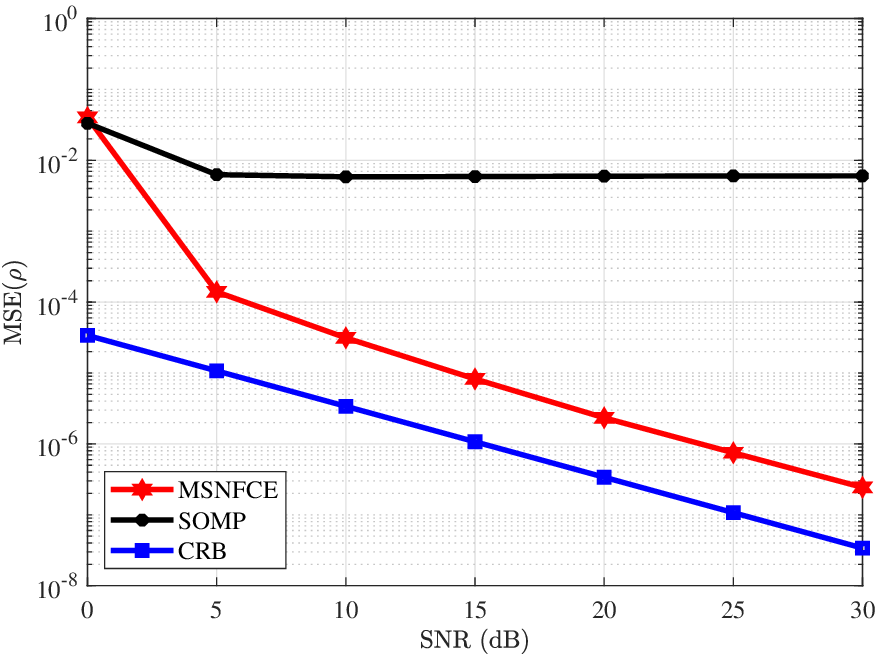}
			\caption*{(e) $\rm MSE$ of $\boldsymbol{\rho}$}
			\label{F1_rho}
		\end{minipage}%
	}%
	\subfigure{
		\begin{minipage}[h]{0.33\linewidth}
			\centering
			\includegraphics[width=2.3in]{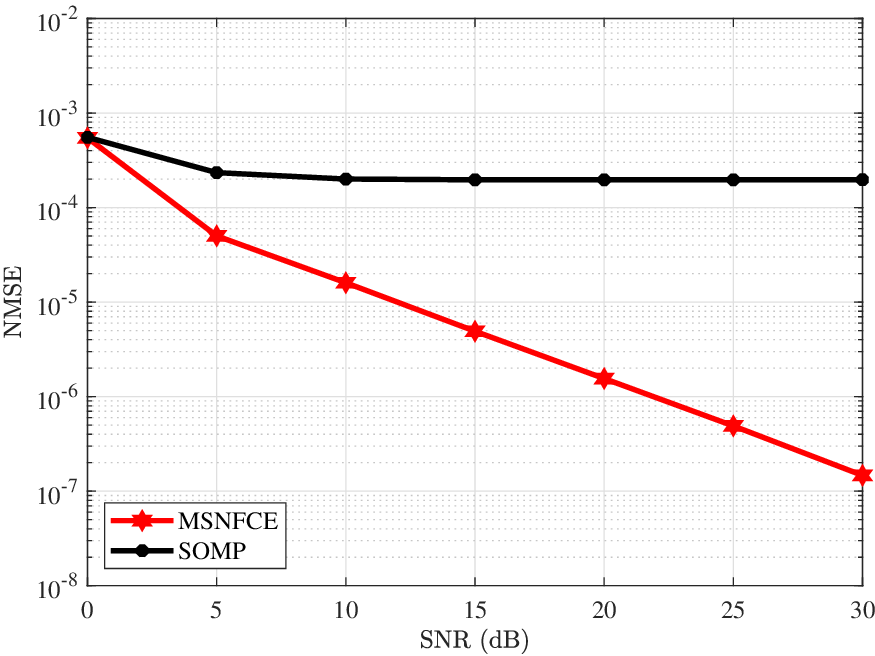}
			\caption*{(f) ${\rm NMSE}$ of cascade channel}
			\label{F1_NMSE}
		\end{minipage}%
	}%
	\centering
	\caption{Performance comparison for channel estimation versus SNR.}
	\label{F1}
\end{figure*}
CRB is commonly utilized to establish a fundamental lower bound on the variance of unbiased estimators for unknown parameters, defining the best possible estimation precision theoretically attainable \cite{kay1993fundamentals}. To evaluate the theoretical performance benchmark for the proposed method, the CRB of the estimated channel parameters $\{ \tilde{\theta}_{\rm RM}^{l}, {\tilde{\phi}}_{\rm RM}^{l}, \tilde{d}_{\rm RM}^{l}, \tilde{\tau}_l,  \tilde{\rho}_l \}_{l=1}^{L}$ is derived. For deducing convenience, we first let $\boldsymbol{\theta}_{\rm RM} = [{\theta}_{\rm RM}^1,\cdots,{\theta}_{\rm RM}^L]^T$, $\boldsymbol{\phi}_{\rm RM} = [{\phi}_{\rm RM}^1, \cdots,{\phi}_{\rm RM}^L]^T$,  $\boldsymbol{\tau} = [\tau_1,\cdots,\tau_L]^T$, $\boldsymbol{d}_{\rm RM} = [d_{\rm RM}^1, \cdots,d_{\rm RM}^L]^T$, $\boldsymbol{\rho}=[\rho_1,\cdots,\rho_L]^T$, and $\boldsymbol{\varsigma} = [\boldsymbol{\theta}_{\rm RM}^T, \boldsymbol{\phi}_{\rm RM}^T, \boldsymbol{\tau}^T, \boldsymbol{d}_{\rm RM}^T, \boldsymbol{\rho}^T]^T$. Then, by performing the mode-$1$ unfolding for the noisy received signal $\boldsymbol{ {\cal Y}}$ and the clean received signal $\boldsymbol{ {\cal G}}$, we can obtain ${\bf Y}_{(1)}$ and ${\bf G}_{(1)}$, respectively. Subsequently, by performing vectoring operator, we can obtain ${\bf y}_{(1)} = {\rm vec}({\bf Y}_{(1)})$ and ${\bf g}_{(1)} = {\rm vec}({\bf G}_{(1)})$. For ${\bf y}_{(1)}$, it can be regard as stacking $\{{\bf y}_{m,k}\}_{m=1:M}^{k=1:K}$ in column, i.e. ${\bf y}_{(1)} = [{\bf y}_{1,1}^T,\cdots,{\bf y}_{m,1}^T,\cdots,{\bf y}_{M,1}^T,\cdots, {\bf y}_{1,k}^T,\cdots,{\bf y}_{m,k}^T,\cdots,{\bf y}_{M,k}^T,$ $\cdots, {\bf y}_{1,K}^T,$ $\cdots,{\bf y}_{m,K}^T,\cdots,{\bf y}_{M,K}^T]^T$, which is the same for ${\bf g}_{(1)}$. Subsequently, the log-likelihood function of $\boldsymbol{\varsigma}$ can be given by
\begin{equation}
\begin{aligned}
{\cal F}(\boldsymbol{\varsigma}) = \hat{d} - ({\bf y}_{(1)} - {\bf g}_{(1)})^H {\bf C}_{\boldsymbol{ {\cal N}} }^{-1} ({\bf y}_{(1)} - {\bf g}_{(1)})
\end{aligned}
\label{log-likelihood_function.eqn}
\end{equation}
where $\hat{d}$ is a constant, and ${\bf C}_{\boldsymbol{ {\cal N}}}$ is the covariance matrix associated with the noise vector ${\rm vec}({\bf N}^{\bf W}_{(1)})$. 

Then, the complex Fisher information matrix (FIM) is given by
\begin{equation}
\begin{aligned}
{\rm FIM}(\boldsymbol{\varsigma}) = {\mathbb E} \left\{ \left(\frac{\partial {\cal F}(\boldsymbol{\varsigma})}{\partial \boldsymbol{\varsigma}}\right)^H  \frac{\partial {\cal F}(\boldsymbol{\varsigma})}{\partial \boldsymbol{\varsigma}} \right\}
\end{aligned}
\label{partial_derivation_log-likelihood_function.eqn}
\end{equation}

Finally, the CRB for $\boldsymbol{\varsigma}$ is derived as
\begin{equation}
\begin{aligned}
{\rm CRB}(\boldsymbol{\varsigma}) = {\rm FIM}^{-1}(\boldsymbol{\varsigma}) 
\end{aligned}
\label{CRB.eqn}
\end{equation}
 
A comprehensive closed-form derivation for FIM is presented in Appendix \ref{CRB_deviation}.

\section{Simulation Results} \label{simulation_results}
To demonstrate the efficacy of the proposed MSNFCE method, numerical experiments are conducted under the following settings. System parameters include a central frequency $f_c = 30$ GHz, bandwidth $f_s = 1$ GHz, and $N_{\rm BS}=N_{\rm UE}=32$ antennas equipped at the BS and UEs. The RIS is a circular array with $N_{\rm R} = 1024$ unit cells and radius $r_c= \lambda/(4{\rm sin}(\pi/N_{\rm R}))$, yielding an inter-element spacing of $\lambda/2$. The number of data streams is $N_s = 4$. Regarding the mmWave channel, a single LoS path connects the BS and RIS, whereas $L=3$ paths link the RIS and UEs. Angular parameters are drawn from the intervals: $\theta_{\rm BR} \in (-\pi, -\pi/2)$, $\phi_{\rm BR} \in (0, \pi/2)$, $\theta_{\rm RM}^l \in (0,\pi)$, and $\phi_{\rm RM}^l \in (-\pi /2,0)$. The complex gains $\alpha$ and $\beta_l$ are drawn from ${\cal CN}(0,1)$. The hybrid precoder $\bf F$ and combiner $\bf W$ are initialized with entries from a unit circle and normalized column-wise to meet power constraints. For the NOMA configuration, there are $N = 256$ subcarriers, with the first $K = 20$ allocated as pilots, and $4$-QAM is adopted for signal modulation. The power allocation factors are set to $\iota_1 = 0.8$ for UE $1$ (low-order user) and $\iota_2 = 0.2$ for UE $2$ (high-order user). Finally, the channel estimation process employs $M=32$ time slots, each containing $N_b=7$ NOMA symbols.
The CRB serves as the benchmark in the simulations, while the SOMP \cite{9760391} is chosen as comparison algorithm for performance evaluation. To the best of our knowledge, the research on near-field channel estimation with beam squint effect is still inadequate. Due to the destruction of low-rank structures, related researches primarily utilize the OMP algorithm with high-dimensional basis for channel estimation, which is impossible to make a good comparison in the case of involving multiple variables. Thus, we refer to the SOMP method in \cite{9760391} to estimate $\boldsymbol{\theta}_{\rm RM}$, $\boldsymbol{\phi}_{\rm RM}$, and $\boldsymbol{d}_{\rm RM}$ with basis after dimensionality reduction. However, owing to the differences in models, we only adopt the core idea of SOMP method for comparison, and the estimation of the remaining parameters are consistent with the proposed method.

For performance analysis purposes, we define the SNR as ${\rm SNR } = \Vert{\boldsymbol{{\cal Y - N}}^{\bf W}}\Vert_F^2/ \Vert{\boldsymbol{{\cal N}}^{\bf W}}\Vert_F^2$, and employ the mean square error (MSE) to assess the estimation accuracy of the channel parameters, expressed as
\begin{equation}
\begin{aligned}
{\rm MSE}({\boldsymbol{\eta} }) = \frac{1}{L} \Vert {\boldsymbol{\eta}} - { \tilde{\boldsymbol{\eta}}} \Vert_2^2
\end{aligned}
\label{MSE.eqn}
\end{equation}
Additionally, the normalized mean square error (NMSE) is adopted for evaluating the estimation performance of the  cascade channel, expressed as
\begin{equation}
\begin{aligned}
{\rm NMSE} = \frac{1}{N} \sum_{n=1}^{N} \frac{ \Vert {\bf O}[k] - \tilde{{\bf O}}[k] \Vert_F^2}{\Vert {\bf O}[k] \Vert_F^2}
\end{aligned}
\label{NMSE.eqn}
\end{equation}
Here, we average all simulation results over independent Monte Carlo trials.

The estimation performance of the channel parameters versus SNR for the discussed methods is illustrated in Fig. \ref{F1}. As shown in Fig. \ref{F1_theta_RM}, the MSE of $\boldsymbol{\theta}_{\rm RM}$ estimated by the proposed method decreases as SNR increases and approaches the CRB most closely. Owing to the refinement-assisted estimator and Vandermonde property of the constructed matrix, the proposed MSNFCE method exhibits stable estimation performance. In contrast, limited by grid mismatch problem, the MSE of the SOMP method first increases and then tends to stabilize with the increasing SNR, reaching performance saturation. Similar MSE results for the remaining channel parameters $\boldsymbol{\phi}_{\rm RM}$, $\boldsymbol{\tau}$, $\boldsymbol{d}$, and $\boldsymbol{\rho}$ are presented in Figs. \ref{F1_phi_RM}-\ref{F1_rho}. To further evaluate the estimation performance of the cascade channel, the NMSE is calculated and shown in Fig. \ref{F1_NMSE}. It can be observed from Fig. \ref{F1_NMSE} that the NMSE of the proposed method decays exponentially with increasing SNR, whereas that of the SOMP method initially decreases but quickly reaches a saturation region. Overall, from Fig. \ref{F1}, it can be confirmed that the propose method achieves high-precision channel estimation.
\begin{figure}[htbp]
	\centering
	\includegraphics[width=3.3 in]{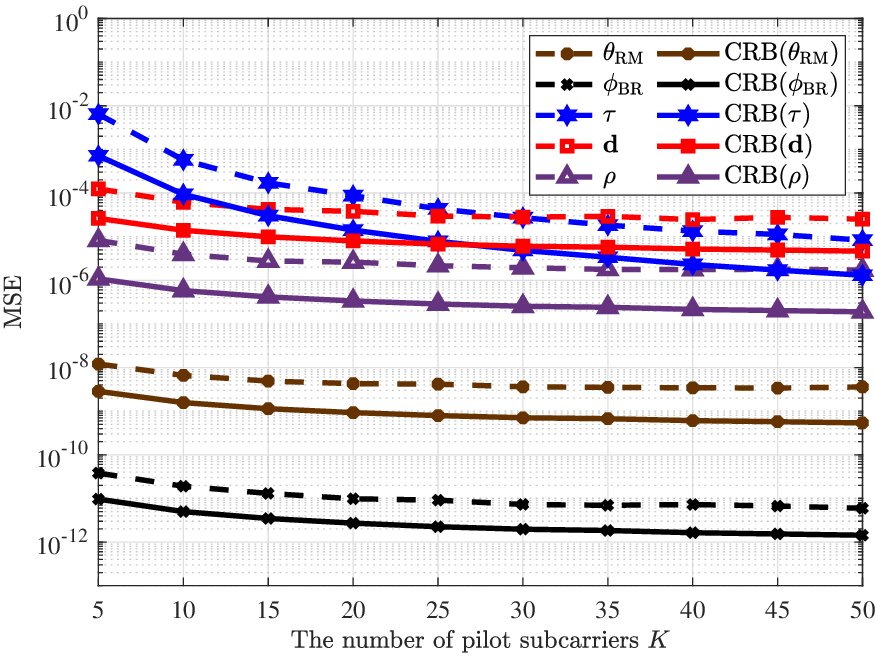}
	\caption{Performance comparison for channel estimation versus the number of pilot subcarriers.}
	\label{versus_pilot}
\end{figure}

Furthermore, To further investigate how the number of pilot subcarriers influences the proposed method, we conduct experiments on the channel estimation performance versus the number of pilot subcarriers at ${\rm SNR} = 20$ dB, with the results illustrated in Fig. \ref{versus_pilot}. It can be observed from Fig. \ref{versus_pilot} that the MSE of the estimated parameters obtained by the proposed method decreases initially and then tends to remain constant as the number of pilot subcarriers increases, whose trend of change is consistent with the CRB. Although a larger number of pilot subcarriers leads to higher estimation performance, it also brings higher computational complexity. Consequently, an appropriate trade-off between estimation accuracy and computational efficiency should be maintained. By comprehensively evaluating diverse performance metrics, this paper establishes $K = 20$ as the baseline configuration for the system.
\begin{figure}[htbp]
	\centering
	\includegraphics[width=3.3 in]{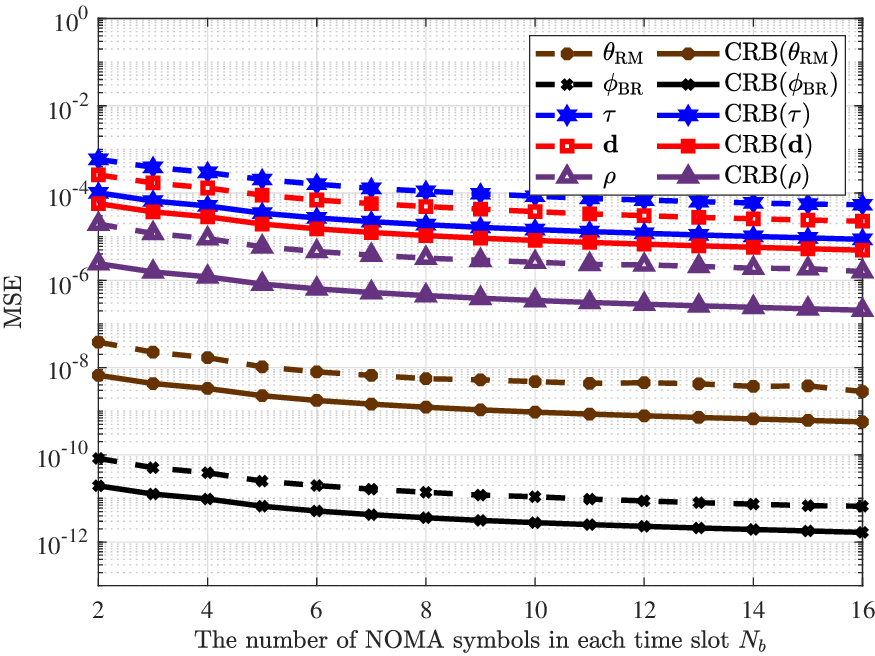}
	\caption{Performance comparison for channel estimation versus the number of NOMA symbols in each time slot.}
	\label{versus_Nb}
\end{figure}

Similarly, Fig. \ref{versus_Nb} shows the channel estimation performance versus the number of NOMA symbols in each time slot when ${\rm SNR} = 20$ dB and $K = 20$. It can be observed from Fig. \ref{versus_Nb} that the MSE of the estimated channel parameters obtained by the proposed method also decreases initially and then tends to remain constant as the number of NOMA symbols in each time slot increases, which has the similar trend of change as the CRB. Accordingly, this paper adopts $N_b = 7$ as the default setting of the system.
\section{Conclusion}
In this paper, we proposed a multi-stage near-field channel estimation method for XL-CRIS-aided mmWave MIMO-NOMA system with beam squint effect. By designing an XL-CRIS architecture with rotational symmetry, we achieved an angle-invariant near-field region, ensuring a constant effective aperture and avoiding hybrid field channel modeling. Then, we formulated the received signal as a third-order tensor, based on which a multi-stage estimation framework was developed to decouple the high-dimensional problem into several subproblems while preserving path-wise parameter pairing through a shared permutation matrix. Finally, theoretical analysis via the derived vector-form Cramér–Rao bound (CRB) and numerical simulation results confirmed the advanced estimation performance of the proposed method. 

\appendices
\section{Deviation of CRB}	\label{CRB_deviation}
As a preliminary step for subsequent analysis, we derive the covariance matrix of the noise vector before computing the CRB for each parameter. Firstly, by applying the mode-$1$ unfolding to the noise tensor $\boldsymbol{{\cal N}}^{\bf W} ={\rm Vec}^1_2(\boldsymbol{{\cal N}} \times_1 {\bf W}^T)$, we have ${\bf N}^{\bf W}_{(1)} = {\bf W}^T  {\bf N}_{(1)}$, where ${\bf N}_{(1)}$ denotes the mode-$1$ unfolding form of $\boldsymbol{{\cal N}}$. From this, we obtain the vectorized form of ${\bf N}^{\bf W}_{(1)}$ as
\begin{equation}
\begin{aligned}
{\rm vec}({\bf N}^{\bf W}_{(1)}) &= {\rm vec}({\bf W}^T  {\bf N}_{(1)}) \\
&= ( {\bf I}_{N_b M K} \otimes {\bf W}^T) {\rm vec}({\bf N}_{(1)})
\end{aligned}
\label{vec_nw.eqn}
\end{equation}

Then, we can calculate second-order moments ${\bf M}_{\boldsymbol{{\cal N}}}$ and covariance matrix ${\bf C}_{\boldsymbol{ {\cal N}}}$ by
\begin{equation}
\begin{aligned}
{\bf M}_{\boldsymbol{{\cal N}}} &= {\mathbb E}\{{\rm vec}({\bf N}^{\bf W}_{(1)}) {\rm vec}({\bf N}^{\bf W}_{(1)})^T \}  \\
&= {\mathbb E}\{{\rm vec}({\bf N}^{\bf W}_{(1)})^* {\rm vec}({\bf N}^{\bf W}_{(1)})^H \} \\
& = {\bf 0} 
\\
{\bf C}_{\boldsymbol{ {\cal N}}} &= {\mathbb E}\{ {\rm vec}({\bf N}^{\bf W}_{(1)}) {\rm vec}({\bf N}^{\bf W}_{(1)})^H \} \\
&=\sigma^2 ({\bf I}_{N_b M K } \otimes ({\bf W}^T {\bf W}^*)) 
\end{aligned}
\label{Cn_calculate.eqn}
\end{equation}

We now provide a step-by-step derivation of the complex FIM. The partial derivative of ${\cal F}(\boldsymbol{\varsigma})$ with respect to ${\theta}_{\rm RM}^l$ is first obtained as
\begin{equation}
\begin{aligned}
\frac{\partial {\cal F}(\boldsymbol{\varsigma})}{\partial {\theta}_{\rm RM}^l}= \left(\frac{\partial {\cal F}(\boldsymbol{\varsigma})}{\partial {\bf g}_{(1)}}\right)^T  \frac{\partial {\bf g}_{(1)}}{\partial {\theta}_{\rm RM}^l} +    \left(\frac{\partial {\cal F}(\boldsymbol{\varsigma})}{\partial {\bf g}_{(1)}^*}\right)^T  \frac{\partial {\bf g}_{(1)}^*}{\partial {\theta}_{\rm RM}^l} 
\end{aligned}
\label{partial_derivation_phi_BR.eqn}
\end{equation}
where 
\begin{equation}
\begin{aligned}
&\frac{\partial {\cal F}(\boldsymbol{\varsigma})}{\partial {\bf g}_{(1)}} = {\bf C}_{\boldsymbol{ {\cal N}}}^{-T}({\bf y}_{(1)} - {\bf g}_{(1)})^* = {\bf C}_{\boldsymbol{ {\cal N}}}^{-T} {\rm vec}({\bf N}^{\bf W}_{(1)})^*\\
& \frac{\partial {\cal F}(\boldsymbol{\varsigma})}{\partial {{\bf g}_{(1)}^*}} =\left (\frac{\partial {\cal F}(\boldsymbol{\varsigma})}{\partial {\bf g}_{(1)}} \right)^* ~~~~ \frac{\partial {\bf g}_{(1)}^*}{\partial {{\theta}_{\rm RM}^l}} =\left (\frac{\partial {\bf g}_{(1)}}{\partial  {\theta}_{\rm RM}^l}\right)^* \\
&\frac{\partial {\bf g}_{(1)}}{\partial {{\theta}_{\rm RM}^l}} = [\frac{\partial{\bf g}_{1,1}^T}{\partial {{\theta}_{\rm RM}^l}},\cdots,\frac{\partial{\bf g}_{m,1}^T}{\partial {{\theta}_{\rm RM}^l}},\cdots,\frac{\partial{\bf g}_{M,1}^T}{\partial {{\theta}_{\rm RM}^l}},\cdots, \frac{\partial{\bf g}_{1,k}^T}{\partial {{\theta}_{\rm RM}^l}},\cdots,\\
&\frac{\partial{\bf g}_{m,k}^T}{\partial {{\theta}_{\rm RM}^l}},\cdots,\frac{\partial{\bf g}_{M,k}^T}{\partial {{\theta}_{\rm RM}^l}},\cdots, \frac{\partial{\bf g}_{1,K}^T}{\partial {{\theta}_{\rm RM}^l}}, \cdots,\frac{\partial{\bf g}_{m,K}^T}{\partial {{\theta}_{\rm RM}^l}},\cdots,\frac{\partial{\bf g}_{M,K}^T}{\partial {{\theta}_{\rm RM}^l}}]^T
\end{aligned}
\label{partial_derivation_phi_BR_where.eqn}
\end{equation}
where 
\begin{equation}
\begin{aligned}
&\frac{\partial{\bf g}_{m,k}}{\partial {{\theta}_{\rm RM}^l}} = \rho_l  e^{-j \frac{2 \pi  f_s}{N}(k-1) \tau_l} (-j \pi f_k / f_c) {\rm sin}({\theta}_{\rm RM}^l)  \boldsymbol{\Upsilon}^T ({\bf I}_{N_{\rm BS}}\\ 
&  ~~~~~~~~~~\otimes
 {\rm diag}([0,1,\cdots, {N}_{\rm MS}-1])) {\bf a}_{s}(\phi_{\rm BR},\theta_{\rm RM}^l,f_k)\\
& ~~~~~~~~~~ {\bf a}_{r}^T(\theta_{\rm BR},\phi_{\rm RM}^l,f_k,d_{\rm BR},d_{\rm RM}^l) \boldsymbol{\Psi}_{m}
\end{aligned}
\label{zmk_where.eqn}
\end{equation}

Therefore, we have
\begin{equation}
\begin{aligned}
\frac{\partial {\cal F}(\boldsymbol{\varsigma})}{\partial {{\theta}_{\rm RM}^l}} &= {\rm vec}({\bf N}^{\bf W}_{(1)})^H  {\bf C}_{\boldsymbol{ {\cal N}}}^{-1} \frac{\partial {\bf g}_{(1)}}{\partial {{\theta}_{\rm RM}^l}} + {\rm vec}({\bf N}^{\bf W}_{(1)})^T  {\bf C}_{\boldsymbol{ {\cal N}}}^{-*} (\frac{\partial {\bf g}_{(1)}}{\partial {{\theta}_{\rm RM}^l}})^* \\
& = 2 {\rm Re} \left\{  {\rm vec}({\bf N}^{\bf W}_{(1)})^H  {\bf C}_{\boldsymbol{ {\cal N}}}^{-1} \frac{\partial {\bf g}_{(1)}}{\partial {{\theta}_{\rm RM}^l}}    \right\}
\end{aligned} 
\label{Lp_theta.eqn}
\end{equation}
where ${\rm Re}\{{\cdot}\}$ extracts the real part of a complex number. 

Likewise, the partial derivatives for the other channel parameters are given by
\begin{equation}
\begin{aligned}
\frac{\partial {\cal F}(\boldsymbol{\varsigma})}{\partial {\phi}_{\rm RM}^l} &= 2 {\rm Re} \left\{  {\rm vec}({\bf N}^{\bf W}_{(1)})^H  {\bf C}_{\boldsymbol{ {\cal N}}}^{-1} \frac{\partial {\bf g}_{(1)}}{\partial {{\phi}_{\rm RM}^l}}  \right\}  \\
\frac{\partial {\cal F}(\boldsymbol{\varsigma})}{\partial {\tau}_l} &= 2 {\rm Re} \left\{  {\rm vec}({\bf N}^{\bf W}_{(1)})^H  {\bf C}_{\boldsymbol{ {\cal N}}}^{-1} \frac{\partial {\bf g}_{(1)}}{\partial {\tau}_l} \right\} \\
\frac{\partial {\cal F}(\boldsymbol{\varsigma})}{\partial d_{\rm RM}^l} &= 2 {\rm Re} \left\{  {\rm vec}({\bf N}^{\bf W}_{(1)})^H  {\bf C}_{\boldsymbol{ {\cal N}}}^{-1} \frac{\partial {\bf g}_{(1)}}{\partial d_{\rm RM}^l} \right\}\\
\frac{\partial {\cal F}(\boldsymbol{\varsigma})}{\partial \rho_l} &=  {\rm vec}({\bf N}^{\bf W}_{(1)})^H  {\bf C}_{\boldsymbol{ {\cal N}}}^{-1} \frac{\partial {\bf g}_{(1)}}{\partial  \rho_l} 
\end{aligned} 
\label{Lp_phi_tau_fl.eqn}
\end{equation}
where
\begin{equation}
\begin{aligned}
&\frac{\partial {\bf g}_{(1)}}{\partial {{\phi}_{\rm RM}^l}}  =  [\frac{\partial{\bf g}_{1,1}^T}{\partial {{\phi}_{\rm RM}^l}},\cdots,\frac{\partial{\bf g}_{M,1}^T}{\partial {{\phi}_{\rm RM}^l}},\cdots, \frac{\partial{\bf g}_{1,k}^T}{\partial {{\phi}_{\rm RM}^l}},\cdots,\frac{\partial{\bf g}_{M,k}^T}{\partial {{\phi}_{\rm RM}^l}},\\
&~~~~~~~~~~~~\cdots, \frac{\partial{\bf g}_{1,K}^T} {\partial {{\phi}_{\rm RM}^l}},\cdots,\frac{\partial{\bf g}_{M,K}^T}{\partial {{\phi}_{\rm RM}^l}}]^T \\
& \frac{\partial {\bf g}_{(1)}}{\partial {\tau}_l} =  [\frac{\partial{\bf g}_{1,1}^T}{\partial {\tau}_l},\cdots,\frac{\partial{\bf g}_{M,1}^T}{\partial {\tau}_l},\cdots, \frac{\partial{\bf g}_{1,k}^T}{\partial {\tau}_l},\cdots,\frac{\partial{\bf g}_{M,k}^T}{\partial {\tau}_l},\\
&~~~~~~~~~~~~\cdots, \frac{\partial{\bf g}_{1,K}^T} {\partial {\tau}_l},\cdots,\frac{\partial{\bf g}_{M,K}^T}{\partial {\tau}_l}]^T \\
& \frac{\partial {\bf g}_{(1)}}{\partial d_{\rm RM}^l}=  [\frac{\partial{\bf g}_{1,1}^T}{\partial d_{\rm RM}^l},\cdots,\frac{\partial{\bf g}_{M,1}^T}{\partial d_{\rm RM}^l},\cdots, \frac{\partial{\bf g}_{1,k}^T}{\partial d_{\rm RM}^l},\cdots,\frac{\partial{\bf g}_{M,k}^T}{\partial d_{\rm RM}^l},\\
&~~~~~~~~~~~~\cdots, \frac{\partial{\bf g}_{1,K}^T} {\partial d_{\rm RM}^l},\cdots,\frac{\partial{\bf g}_{M,K}^T}{\partial d_{\rm RM}^l}]^T \\
& \frac{\partial {\bf g}_{(1)}}{\partial  \rho_l}  =  [\frac{\partial{\bf g}_{1,1}^T}{\partial  \rho_l},\cdots,\frac{\partial{\bf g}_{M,1}^T}{\partial  \rho_l},\cdots, \frac{\partial{\bf g}_{1,k}^T}{\partial  \rho_l},\cdots,\frac{\partial{\bf g}_{M,k}^T}{\partial  \rho_l},\\
&~~~~~~~~~~~~\cdots, \frac{\partial{\bf g}_{1,K}^T} {\partial  \rho_l},\cdots,\frac{\partial{\bf g}_{M,K}^T}{\partial  \rho_l}]^T \\
&\frac{\partial{\bf g}_{m,k}}{\partial {{\phi}_{\rm RM}^l}} = \rho_l  e^{-j \frac{2 \pi  f_s}{N}(k-1) \tau_l}   \boldsymbol{\Upsilon}^T  {\bf a}_{s}(\phi_{\rm BR},\theta_{\rm RM}^l,f_k)\\
& ~~~~~~~~~~ {\bf a}_{r}^T(\theta_{\rm BR},\phi_{\rm RM}^l,f_k,d_{\rm BR},d_{\rm RM}^l)(j 2 \pi f_k  \frac{r_c}{v_c} \\
&~~~~~~~~~~( -{\rm sin}(\phi_{\rm RM}^l - {\boldsymbol{\zeta}}) - r_c / d_{\rm RM}^l {\rm sin}(\phi_{\rm RM}^l - {\boldsymbol{\zeta}} ) \circledast\\
&~~~~~~~~~~ {\rm cos}(\phi_{\rm RM}^l - {\boldsymbol{\zeta}})) \circledast \boldsymbol{\Psi}_{m})\\
&\frac{\partial{\bf g}_{m,k}}{\partial {\tau}_l} = \rho_l  (-j \frac{2 \pi  f_s}{N}(k-1)) e^{-j \frac{2 \pi  f_s}{N}(k-1) \tau_l}   \boldsymbol{\Upsilon}^T \\ 
&  ~~~~~~ {\bf a}_{s}(\phi_{\rm BR},\theta_{\rm RM}^l,f_k) {\bf a}_{r}^T(\theta_{\rm BR},\phi_{\rm RM}^l,f_k,d_{\rm BR},d_{\rm RM}^l) \boldsymbol{\Psi}_{m}\\
&\frac{\partial{\bf g}_{m,k}}{\partial d_{\rm RM}^l} = \rho_l  e^{-j \frac{2 \pi  f_s}{N}(k-1) \tau_l}  \boldsymbol{\Upsilon}^T {\bf a}_{s}(\phi_{\rm BR},\theta_{\rm RM}^l,f_k)\\
& ~~~~~~~~~~ {\bf a}_{r}^T(\theta_{\rm BR},\phi_{\rm RM}^l,f_k,d_{\rm BR},d_{\rm RM}^l)(j 2 \pi f_k  \frac{r_c}{v_c} ( \frac{r_c}{2 (d_{\rm RM}^l)^2}  \\
&~~~~~~~~~~ ( {\rm sin}(\phi_{\rm RM}^l - {\boldsymbol{\zeta}})\circledast {\rm sin}(\phi_{\rm RM}^l - {\boldsymbol{\zeta}}) ) \circledast \boldsymbol{\Psi}_{m})\\
\end{aligned} 
\label{Lp_phi_tau_fl_partial.eqn}
\end{equation}
\begin{equation*}
\begin{aligned}
&\frac{\partial{\bf g}_{m,k}}{\partial  \rho_l} = e^{-j \frac{2 \pi  f_s}{N}(k-1) \tau_l} \boldsymbol{\Upsilon}^T {\bf a}_{s}(\phi_{\rm BR},\theta_{\rm RM}^l,f_k)\\
& ~~~~~~~~~~ {\bf a}_{r}^T(\theta_{\rm BR},\phi_{\rm RM}^l,f_k,d_{\rm BR},d_{\rm RM}^l) \boldsymbol{\Psi}_{m} 
\end{aligned} 
\label{Lp_phi_tau_fl_partial_where.eqn}
\end{equation*}

Next, we evaluate the entries comprising the principal minors of ${\cal F}(\boldsymbol{\varsigma})$. Specifically, the $(l_1,l_2)$-th entry of ${\mathbb E} \left\{ \left(\frac{\partial {\cal F}(\boldsymbol{\varsigma})}{\partial \boldsymbol{\theta}_{\rm RM}}\right)^H  \frac{\partial {\cal F}(\boldsymbol{\varsigma})}{\partial \boldsymbol{\theta}_{\rm RM}} \right\}$ is given by
\begin{equation}
\begin{aligned}
&{\mathbb E} \left\{ \left(\frac{\partial {\cal F}(\boldsymbol{\varsigma})}{\partial  \theta_{\rm RM}^{l_1}} \right)^*  \frac{\partial {\cal F}(\boldsymbol{\varsigma})}{\partial  \theta_{\rm RM}^{l_2}} \right\} = \\
&{\mathbb E}   \left\{ \begin{aligned}
({\rm vec}({\bf N}^{\bf W}_{(1)})^H  {\bf C}_{\boldsymbol{ {\cal N}}}^{-1} \frac{\partial {\bf g}_{(1)}}{\partial {{\theta}_{\rm RM}^{l_1}}} +  ({\rm vec}({\bf N}^{\bf W}_{(1)})^H  {\bf C}_{\boldsymbol{ {\cal N}}}^{-1} \frac{\partial {\bf g}_{(1)}}{\partial {{\theta}_{\rm RM}^{l_1}}})^*)^T \\
({\rm vec}({\bf N}^{\bf W}_{(1)})^H  {\bf C}_{\boldsymbol{ {\cal N}}}^{-1} \frac{\partial {\bf g}_{(1)}}{\partial {{\theta}_{\rm RM}^{l_2}}} +  ({\rm vec}({\bf N}^{\bf W}_{(1)})^H  {\bf C}_{\boldsymbol{ {\cal N}}}^{-1} \frac{\partial {\bf g}_{(1)}}{\partial {{\theta}_{\rm RM}^{l_2}}})^*)
\end{aligned}
\right \} \\
&=\left( \frac{\partial {\bf g}_{(1)}}{\partial {{\theta}_{\rm RM}^{l_1}}}\right)^T {\bf C}_{\boldsymbol{ {\cal N}}}^{-T} \left( \frac{\partial {\bf g}_{(1)}}{\partial {{\theta}_{\rm RM}^{l_2}}}\right)^*+\left( \frac{\partial {\bf g}_{(1)}}{\partial {{\theta}_{\rm RM}^{l_1}}}\right)^H {\bf C}_{\boldsymbol{ {\cal N}}}^{-1}  \frac{\partial {\bf g}_{(1)}}{\partial {{\theta}_{\rm RM}^{l_2}}} \\
& = 2{\rm Re} \left\{     \left( \frac{\partial {\bf g}_{(1)}}{\partial {{\theta}_{\rm RM}^{l_1}}}\right)^H {\bf C}_{\boldsymbol{ {\cal N}}}^{-1}  \frac{\partial {\bf g}_{(1)}}{\partial {{\theta}_{\rm RM}^{l_2}}}     \right\}
\end{aligned}
\label{partial_derivation_thetal1l2.eqn}
\end{equation}

We now turn to the off-principal minors of ${\cal F}(\boldsymbol{\varsigma})$. As a representative case, the $(l_1, l_2)$-th entry of ${\mathbb E} \left\{ \left(\frac{\partial {\cal F}(\boldsymbol{\varsigma})}{\partial \boldsymbol{\theta}_{\rm RM}}\right)^H  \frac{\partial {\cal F}(\boldsymbol{\varsigma})}{\partial \boldsymbol{\phi}_{\rm RM}} \right\}$ is given by
\begin{equation}
\begin{aligned}
&{\mathbb E} \left\{ \left(\frac{\partial {\cal F}(\boldsymbol{\varsigma})}{\partial  \theta_{\rm RM}^{l_1}} \right)^*  \frac{\partial {\cal F}(\boldsymbol{\varsigma})}{\partial  \phi_{\rm RM}^{l_2}} \right\} = 2{\rm Re} \left\{     \left( \frac{\partial {\bf g}_{(1)}}{\partial {{\theta}_{\rm RM}^{l_1}}}\right)^H {\bf C}_{\boldsymbol{ {\cal N}}}^{-1}  \frac{\partial {\bf g}_{(1)}}{\partial {{\phi}_{\rm RM}^{l_2}}}     \right\}
\end{aligned}
\label{partial_derivation_thetal1phil2.eqn}
\end{equation}

Following a similar procedure, we obtain
\begin{equation}
\begin{aligned}
&{\mathbb E} \left\{ \left(\frac{\partial {\cal F}(\boldsymbol{\varsigma})}{\partial  {\kappa}_{l_1}} \right)^*  \frac{\partial {\cal F}(\boldsymbol{\varsigma})}{\partial  {\upsilon}_{l_2}} \right\}  = 2{\rm Re} \left\{     \left( \frac{\partial {\bf g}_{(1)}}{\partial {\kappa_{l_1}}}\right)^H {\bf C}_{\boldsymbol{ {\cal N}}}^{-1}  \frac{\partial {\bf g}_{(1)}}{\partial {\upsilon_{l_2}}}     \right\} \\
&{\mathbb E} \left\{ \left(\frac{\partial {\cal F}(\boldsymbol{\varsigma})}{\partial {\rho}_{l_1}} \right)^*  \frac{\partial {\cal F}(\boldsymbol{\varsigma})}{\partial  {\kappa}_{l_2}} \right\}  =     \left( \frac{\partial {\bf g}_{(1)}}{\partial {{\rho}_{l_1}}}\right)^H {\bf C}_{\boldsymbol{ {\cal N}}}^{-1}  \frac{\partial {\bf g}_{(1)}}{\partial { {\kappa}_{l_2}}}   \\
&{\mathbb E} \left\{ \left(\frac{\partial {\cal F}(\boldsymbol{\varsigma})}{\partial \kappa_{l_1}} \right)^*  \frac{\partial {\cal F}(\boldsymbol{\varsigma})}{\partial  \rho_{l_2}} \right\}  = \left( \frac{\partial {\bf g}_{(1)}}{\partial {\kappa_{l_1}}}\right)^H {\bf C}_{\boldsymbol{ {\cal N}}}^{-1}  \frac{\partial {\bf g}_{(1)}}{\partial {\rho_{l_2}}} 
\end{aligned}
\label{partial_derivation_thetaphitau.eqn}
\end{equation}
where $\kappa$ and $\upsilon$ each denote an arbitrary element of $\{{\theta}_{\rm RM}, {\phi}_{\rm RM}, {\tau_l}, d_{\rm RM}, {\rho} \}$.

The derivation of the FIM now completes, allowing the CRB for $\boldsymbol{\varsigma}$ to be calculated from \eqref{CRB.eqn}.
\vspace{-0.2em}
\bibliography{IEEEabrv,reference}
\vfill

\end{document}